\documentclass[%
 reprint,
superscriptaddress,
 amsmath,amssymb,
 aps,
pra,
]{revtex4-2}

\usepackage{graphicx}
\usepackage{dcolumn}
\usepackage{bm}
\usepackage{hyperref}

\usepackage{color}
\usepackage[dvipsnames]{xcolor}
\usepackage{ragged2e}
\usepackage[inkscapelatex=false]{svg}

\begin{document}

\preprint{APS/123-QED}

\title{Electrically Tunable Picosecond-scale  Octupole Fluctuations in Chiral Antiferromagnets}

\author{Shiva T.~Konakanchi}
 \email{skonakan@purdue.edu}
\affiliation{Department of Physics and Astronomy, Purdue University, West Lafayette, Indiana 47907, USA}

\author{Sagnik Banerjee}
\affiliation{Elmore Family School of Electrical and
Computer Engineering, Purdue University, West Lafayette,
Indiana 47907, USA}

\author{Mohammad M.~Rahman}
\affiliation{Department of Chemistry and Biochemistry, University of Maryland, College Park, Maryland 20742, USA}

\author{Yuta Yamane}
\affiliation{Research Institute of Electrical Communication, Tohoku University, 2-1-1 Katahira, Aoba-ku, Sendai 980-8577 Japan}
\affiliation{Frontier Research Institute for Interdisciplinary Sciences, Tohoku University, Sendai 980-8578, Japan}

\author{Shun Kanai}
\affiliation{Research Institute of Electrical Communication, Tohoku University, 2-1-1 Katahira, Aoba-ku, Sendai 980-8577 Japan}
\affiliation{Frontier Research Institute for Interdisciplinary Sciences, Tohoku University, Sendai 980-8578, Japan}
\affiliation{WPI-Advanced Institute for Materials Research (WPI-AIMR), Tohoku University, 2-1-1 Katahira, Aoba-ku, Sendai 980–8577, Japan}
\affiliation{Center for Science and Innovation in Spintronics, Tohoku University, 2-1-1 Katahira, Aoba-ku, Sendai 980–8577, Japan}
\affiliation{PRESTO, Japan Science and Technology Agency (JST), Kawaguchi 332-0012, Japan}
\affiliation{Division for the Establishment of Frontier Sciences of Organization for Advanced Studies at Tohoku University, Tohoku University, Sendai 980-8577, Japan}

\author{Shunsuke Fukami}
\affiliation{Research Institute of Electrical Communication, Tohoku University, 2-1-1 Katahira, Aoba-ku, Sendai 980-8577 Japan}
\affiliation{WPI-Advanced Institute for Materials Research, Tohoku University, 2-1-1 Katahira, Aoba-ku, Sendai 980–8577, Japan}
\affiliation{Center for Science and Innovation in Spintronics, Tohoku University, 2-1-1 Katahira, Aoba-ku, Sendai 980–8577, Japan}
\affiliation{Center for Innovative Integrated Electronic Systems, Tohoku University, 468-1 Aramaki Aza Aoba, Aoba-ku, Sendai 980-0845, Japan}
\affiliation{Inamori Research Institute for Science, Kyoto 600-8411, Japan}

\author{Pramey Upadhyaya}
 \email{prameyup@purdue.edu}
\affiliation{Elmore Family School of Electrical and
Computer Engineering, Purdue University, West Lafayette,
Indiana 47907, USA}




\date{\today}

\begin{abstract}
 We present a theory for the relaxation time of the octupole order parameter in nanoscale chiral antiferromagnets (AFMs) coupled to thermal baths and spin injection sources. Using stochastic spin dynamics simulations, we demonstrate that the octupole moment relaxes through two distinct mechanisms---escape over a barrier and precessional dephasing---as the barrier for octupole fluctuations is lowered relative to the thermal energy. Notably, the octupole moment relaxes orders of magnitude faster than the typical dipolar order parameters, reaching picosecond timescales. By combining Langer's theory with an effective low-energy description of octupole dynamics in chiral AFMs, we derive analytical expressions for the relaxation times. We find that relaxation in chiral AFMs parallels dipole relaxation in XY magnets, with exchange fields serving the role of the dipole fields. Further, by drawing on the analogy between order parameter dynamics in XY magnets under spin injection and current-biased Josephson junctions, we propose a new scheme for electrically tuning the octupole relaxation times. Our work offers fundamental insights for the development of next-generation spintronic devices that harness octupole order parameters for information encoding, especially in octupole-based probabilistic computing.
\end{abstract}

\maketitle


\textit{Introduction.}|Spin-split antiferromagnets, which host an order parameter yielding both a negligible net dipole moment and a finite spin splitting of electronic bands at the Fermi level, have recently gained prominence in condensed matter and spintronics research \cite{PhysRevX.12.040501, PhysRevX.12.031042, dal_din_antiferromagnetic_2024}. Fundamentally, these materials provide a model system for studying topological phenomena and exploring novel magnetic phases \cite{smejkal_anomalous_2022, tan_revealing_2024, legrand_room-temperature_2020}. From an applied perspective, they combine the advantages of efficient electrical readout, typical of ferromagnets, with the low stray-field interactions and high-speed spin dynamics of traditional antiferromagnets (AFMs). As a result, spin-split antiferromagnets hold promise for the development of next-generation spintronic devices \cite{naka_spin_2019, feng_anomalous_2022, shao_spin-neutral_2021, ma_multifunctional_2021}.

The Mn$_3$X family of chiral antiferromagnets represents a particularly intriguing class of spin-split AFMs. In these materials, chiral magnetic order with a negligible dipole moment can be characterized by an octupole moment \cite{suzuki_cluster_2017, dong_tunneling_2022}. Experimental demonstrations of both electrical readout—via the anomalous Hall effect \cite{nakatsuji_large_2015, nayak_large_nodate, kiyohara_giant_2016-1} and tunneling magnetoresistance (TMR) \cite{chen_octupole-driven_2023, qin_room-temperature_2023, chou_large_2024}—and electrical manipulations of octupoles —via spin-orbit torque \cite{tsai_electrical_2020, takeuchi_chiral-spin_2021}— has set the stage for constructing high-performance spintronic devices that encode information within the octupole order parameter.

A central question in octupole-moment-based spintronics is how long the information encoded in octupole order remains correlated before being washed out by coupling to thermal baths. Addressing this question is crucial for applications ranging from nonvolatile memory \cite{yang_two-dimensional_2022, nguyen_recent_2024, 9427163} to probabilistic computing \cite{borders_integer_2019, 10.1063/5.0067927, 10068500}. This question is especially relevant for the family of (anti) chiral antiferromagnets (X = Sn, Ge, Ga, etc.), where the competition among magnetic interactions that produces the (anti) chiral order results in relatively low intrinsic energy barriers (compared to $kT$, where $k$ is the Boltzmann constant and $T$ is the bath temperature) \cite{liu_anomalous_2017}. However, to the best of our knowledge, a fundamental understanding of thermal relaxation of octupole order is currently lacking.

In this work, we theoretically investigate the mechanisms and relaxation timescales of the octupole order parameter in nanomagnets of (anti)chiral AFMs coupled to thermal baths and spin injection sources. Using spin dynamics simulations, we show that depending on the energy barrier encountered by thermal fluctuations in flipping the octupoles, the correlations decay via two distinct mechanisms. In the high-barrier regime, thermal kicks from the bath cause relaxation through random flips of the octupoles over the barrier, with the relaxation timescale governed by the mean escape time over the barrier. In contrast, in the low-barrier regime, relaxation occurs via a precessional dephasing mechanism, where thermal fluctuations induce random precessions of octupoles around exchange fields.

Leveraging the hierarchy of energy scales inherent to chiral AFMs, we construct an effective low-energy theory for fluctuating octupole dynamics, providing an analytical understanding of the octupole relaxation times in both high- and low-barrier limits. We find that the octupole relaxation in chiral AFMs can be mapped onto the relaxation of dipoles in XY magnets (i.e.~magnets with strong easy-plane anisotropy and only weak axial anisotropy within the easy-plane). We show that the exchange fields in chiral AFM play the role of dipolar fields in the XY magnets. Due to the significantly larger strength of the exchange fields compared to the typical dipole fields, octupole fluctuation rates are significantly enhanced over their dipolar counterparts, with the relaxation times reaching picosecond scales for the lowest barriers. Inspired by this mapping, and drawing an analogy between the spin dynamics in XY magnets under spin injection and the charge dynamics in Josephson junctions under current bias, we additionally propose and demonstrate a novel mechanism to electrically tune the octupole relaxation times in chiral AFMs by orders of magnitude via spin orbit torques.

Besides revealing underlying relaxation mechanisms, our results offer insights into designing next generation of spintronic devices that encode information in the octupole order parameter. Specifically, they suggest that nanoscale chiral AFMs are promising candidates for applications requiring fast and robust on-chip generation of electrically tunable random numbers, such as probabilistic computing \cite{PhysRevApplied.12.034061, aadit_massively_2022, chowdhury_accelerated_2023}.

\begin{figure}[t]
    \centering
    \includegraphics[width=\columnwidth]{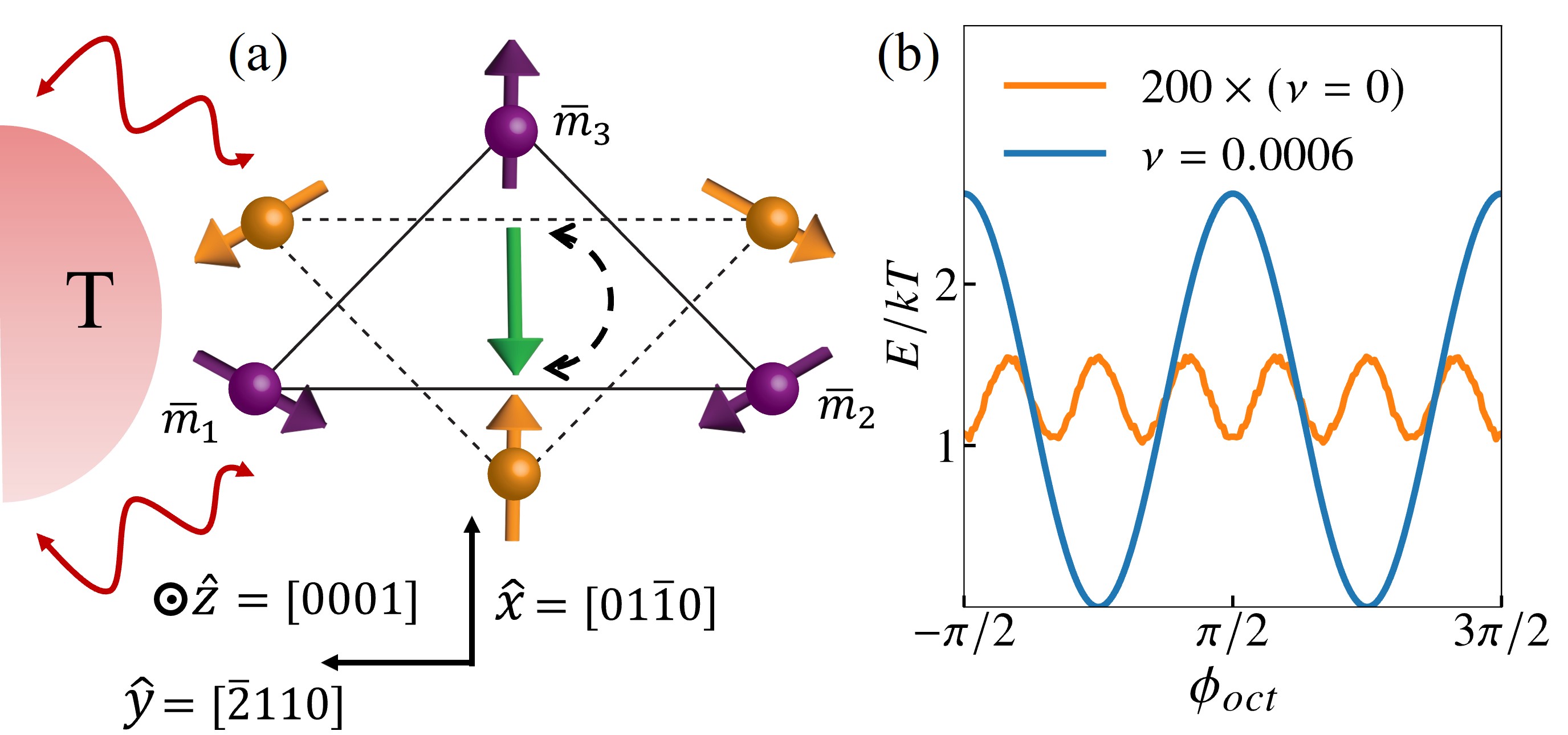}
    \caption{(a) Schematic of the unit cell of Mn$_3$Sn strained along the $y$ axis. The green octupole moment has an easy axis perpendicular to the direction of strain. (b) Energy landscape of a $\sim10^5$ nm$^3$ cylindrical dot as a function of the azimuthal angle, depicting a six-fold ground state ($\nu=0$, scaled by $200$) in bulk and a two-fold ground state in strained thin film limit ($\nu=0.0006$).}
    \label{fig:1}
\end{figure}

\textit{Model} \& \textit {numerics}.|
Mn$_3$X has a layered kagome spin system. Two interpenetrating ($0001$) kagome planes (one below the other) of Mn spins give rise to a unit cell consisting of six spins arranged in a star motif as shown in Fig.~\ref{fig:1}(a). The  intra- and interlayer exchanges between spins are the dominant interactions in the system's Hamiltonian. The former favors a non-collinear alignment among the sublattice spins within each kagome plane, while the latter ensures that the Mn spins in one kagome plane are ferromagnetically aligned with their inverted neighbors in the adjacent plane (see Fig.~\ref{fig:1}(a)) \cite{tomiyoshi_magnetic_1982}. 
Here, we are interested in the stochastic spin dynamics of single domain nanomagnets in the low-energy and long-time limit (as compared to the scales set by the exchange interactions). To this end, we describe the spin state of Mn$_3$X in the micromagnetic continuum limit using three classical magnetization vectors, $\vec{M}_i = M_s \vec{m}_i$, each corresponding to a pair of ferromagnetically locked sublattice spins, with $M_s$ being the saturation magnetization \cite{yamane_dynamics_2019-1, shukla_spin-torque-driven_2022} \footnote{We additionally verify the validity of the three spin model by numerically comparing it against the full six spin unit cell thermal dynamics in the supplement \cite{supplementary}.}.

The free energy density of an Mn$_3$X nanomagnet is then expressed as \cite{takeuchi_chiral-spin_2021, higo_perpendicular_2022, yoon_handedness_2023, miwa_giant_2021}:
\begin{equation}
\mathcal{F} = \sum_{i \neq j} \left[J_{ij} \vec{m}_i \cdot \vec{m}_j + D_{ij} \hat{z} \cdot (\vec{m}_i \times \vec{m}_j)\right]
 - \sum_{i} K_i \left( \vec{m}_i \cdot \vec{e}_i \right)^2.
\label{eqn:hamiltonian}
\end{equation}
Here, $J_{ij}>D_{ij}\gg K_i$ parameterize, respectively, the exchange, Dzyaloshinskii–Moriya (DM), and uniaxial anisotropy energies; $\vec{e}_i$ is the unit vector oriented along the easy axis for $\vec{m}_i$ \cite{supplementary}. For intrinsic single crystals, the hexagonal symmetry dictates that $J_{ij}=J$, $D_{ij}=D$, and $K_i=K$, $\forall \,i,j$. In thin films grown on a substrate, the hexagonal symmetry can be broken, for e.g. via strain due to lattice mismatch \cite{higo_perpendicular_2022}. In the spirit of constructing a minimal model that allows for such asymmetry, we introduce a symmetry breaking parameter $\nu \ll 1$ such that $J_{12}=J(1-\nu)$ \cite{higo_perpendicular_2022, he_magnetic_2024}, while keeping all other interactions to their intrinsic values. 

 Fig.~\ref{fig:1}(a) shows one of the equilibrium magnetic configurations that minimizes $\mathcal{F}$ \cite{higo_perpendicular_2022}: $J$ prefers 120$^0$ arrangement between the moments, while $D$ enforces them to lie in the (easy) $xy$ plane. Additionally, the sign of $D$ fixes the chirality of spin rotations around the unit cell. The case of anti-chiral magnet, i.e. $D>0$, is shown. Weak perturbations due to $\nu$ and $K$, respectively, create an effective uniaxial anisotropy (with the easy axis orthogonal to the symmetry breaking direction) \cite{higo_perpendicular_2022} and slight distortions of the perfect net 120$^0$ alignment \cite{liu_anomalous_2017}. 
 Despite having a negligible net dipole moment, such configurations have experimentally shown significant AHE and TMR effects as a result of the magnetic space group of Mn$_3$X allowing for spin split bands at the Fermi level \cite{liu_anomalous_2017,dong_tunneling_2022}. This splitting, and hence the associated electrical signatures, correspond to ferroic ordering of the so-called octupole moment \cite{suzuki_cluster_2017}. For the case when $\vec{m}_i$ lie in the easy plane, the octupole moment orients along the unit vector:
\begin{equation}
    \vec{m} =M_{xz} [\left( \vec{m}_3 + R \vec{m}_1 + R^2 \vec{m}_2 \right)/3],
    \label{eqn:octupole}
\end{equation}
where $M_{xz}$ is a reflection operator in the $xz$ plane and $R$ represents anticlockwise rotation by $2\pi/3$ around the $z$ axis \cite{go_noncollinear_2022, gomonaj_phenomenologic_1992}. This has raised the intriguing possibility of constructing next generation antiferromagnetic spintronics devices which store information in the octupole moment. 

Coupling Mn$_3$X nanomagnets (of volume $V$) to a thermal bath erases the information encoded in the octupole moments. Fig.~\ref{fig:1}(b) shows typical energy barriers ($\Delta$) that thermal fluctuations must overcome to rotate the octupole moment within the easy plane for intrinsic ($\nu=0$) and symmetry broken ($\nu \neq 0$) case. In the intrinsic case, the competition between exchange, DMI and anisotropy energies gives rise to $\Delta\sim VK^3/J^2 \ll kT$ for typical nanomagnet volumes, effectively creating an XY magnet \cite{liu_anomalous_2017}. In thin films, the energy barriers are thus dominated by extrinsic symmetry breaking, which favors octupole moment to lie along the x-axis with $\Delta\sim2\nu KV$ \cite{supplementary, he_magnetic_2024}. To address the key metric of interest to this work---octupole relaxation times---we describe the fluctuating spin dynamics of an Mn$_3$X nanomagnet coupled to a thermal bath within the stochastic Landau Lifshitz Gilbert (s-LLG) phenomenology by \cite{brown_thermal_1963}: 
\begin{equation}
     \partial_t \vec{m}_i=-\gamma \vec{m}_i \times (-\delta_{M_s\vec{m}_i} \mathcal{F}+ \vec{H}^{\rm th}_i) + \alpha \vec{m}_i \times \partial_t \vec{m}_i.
     \label{sLLG}
\end{equation}Here, $\gamma>0$ and $\alpha$ are the gyromagnetic ratio and the Gilbert damping constant, respectively. $\vec{H}^{\rm th}_i$ are thermal fields that satisfy the fluctuation dissipation theorem \cite{supplementary, brown_thermal_1963}.

\begin{figure*}[t]
    \centering
    \includegraphics[width=\textwidth]{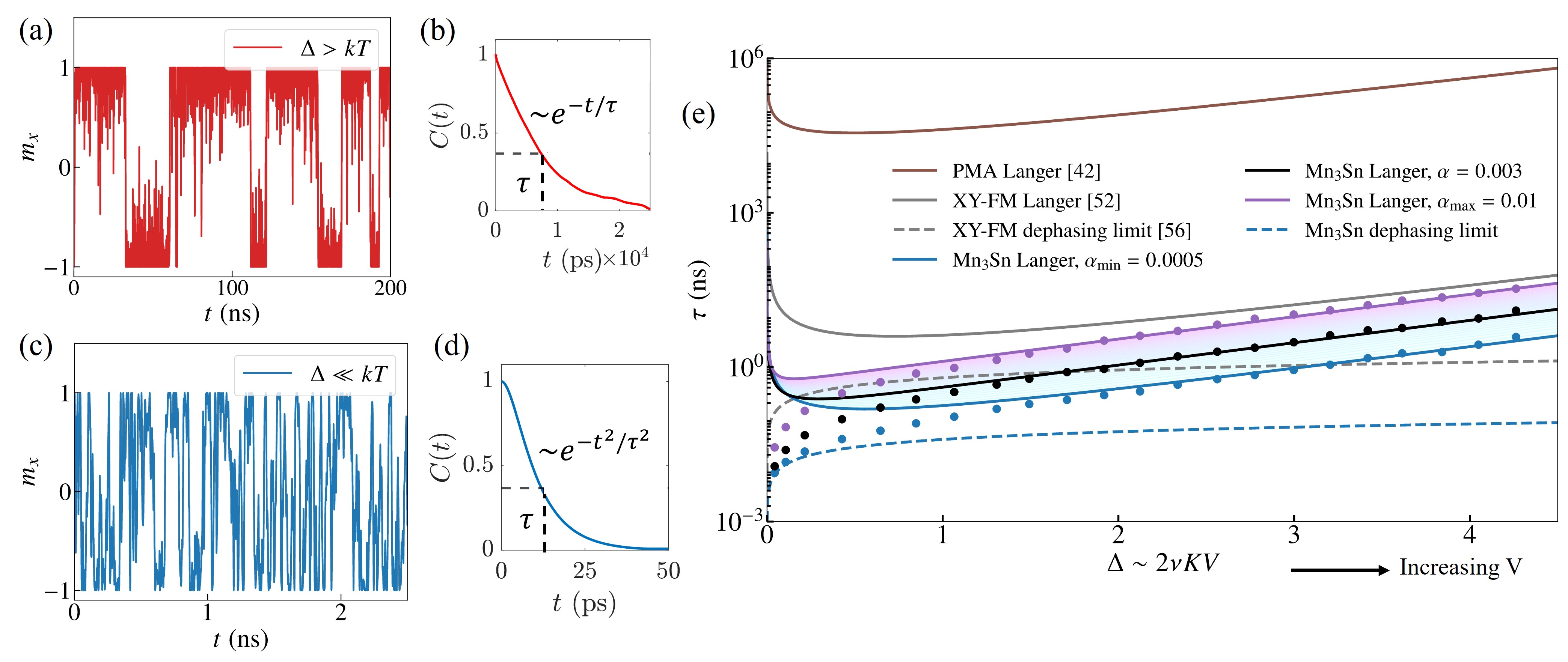}
    \caption{(a) Thermal fluctuations of Mn$_3$Sn octupole moment along its easy axis in the high-barrier limit of $\Delta=4.5 \,kT$. The signal is RTN-like. (b) Autocorrelation function of the time signal from Fig.~\ref{fig:2}(a) showing a decaying exponential trend characteristic of RTN noise. (c) Fluctuations of the octupole moment in the low-barrier limit of $\Delta=0.05 \,kT$. The fluctuations are not RTN-like. (d) Autocorrelation function of the low-barrier octupole fluctuations showing a $\sim e^{-t^2}$ trend, further distinguishing it from the RTN-like signal. (e) Relaxation times for a range of Gilbert damping constants (reported in the literature) and thermal barriers. Solid markers show the results from the numerical simulations and the lines show fits from the analytical formulae derived in this paper. Relaxation times for typical CoFeB ferromagnets with perpendicular magnetic anisotropy (PMA) as well as biaxial anisotropy magnets with strong XY easy-plane and weaker within XY-plane  anisotropy (XY-FM) are also shown for comparison.}
    \label{fig:2}
\end{figure*}

We begin by studying the relaxation time of octupole fluctuations by numerically integrating Eq.~(\ref{sLLG}) and extracting the octupole moment's dynamics of interest by using Eq.~(\ref{eqn:octupole}). By varying the volume, we simulate thermal dynamics of the system in two distinct regimes: $\Delta\ll kT$ (low-barrier limit) and $\Delta> kT$ (high-barrier limit). For numerical arguments throughout the paper, we use Mn$_3$Sn as a model anti-chiral candidate from the Mn$_3$X family. Further details about the numerical simulations and the parameters are provided in the supplementary information \cite{supplementary}.

Figs.~\ref{fig:2}(a) and \ref{fig:2}(c) show time domain fluctuations of the octupole moment in the high-barrier and the low-barrier limits, respectively. As expected, the octupole moment fluctuates much faster in the low-barrier regime. However, we find that while octupole fluctuations in the high-barrier limit are akin to random telgraph noise (RTN), fluctuations in the low-barrier limit do not seem RTN-like. This fact is further highlighted by plotting the octupole autocorrelation functions, $C(t)=\langle m_x(0) m_x(t) \rangle$, in the high and the low-barrier limits. Fig.~\ref{fig:2}(b) shows that the autocorrelation function of octupole fluctuations in the high-barrier limit decays: $C(t)\sim e^{-t/\tau}$. On the other hand, Fig.~\ref{fig:2}(d) shows that the autocorrelation function in the low-barrier limit decays as $\sim e^{\mathcal-t^2/\tau^2}$. As the first main result of this section, this strongly suggests that different physical mechanisms are responsible for relaxation in the two distinct barrier limits. 

To quantify the fluctuation speeds, we define relaxation time, $\tau$, defined as the time taken for the autocorrelation function to fall to $1/e$ of its initial value. 
Fig.~\ref{fig:2}(e) shows the relaxation times for a range of barriers and Gilbert damping constants, whose reported value in the literature falls within the shaded region shown in the figure \cite{tsai_electrical_2020, kanai_theory_2021, yoon_handedness_2023, takeuchi_chiral-spin_2021, zhang_strong_2017}. As a comparison, we also show on the same figure the relaxation times of dipole fluctuations in ferromagnets used in present-day CoFeB-based spintronic devices, i.e. magnets with perpendicular anisotropy (PMA) and biaxial anisotropy magnets with strong XY-easy plane anisotropy and a weaker within XY-plane anisotropy (referred here as XY-FM) \cite{brown_thermal_1963, coffey_thermally_1998, coffey_low_1999, braun_kramerss_1994, coffey_constant-magnetic-field_1995}. 
As the second main result of this section, we note that for all barriers studied here, octupole fluctuations in chiral antiferromagnets are at least an order of magnitude faster than dipole fluctuations in ferromagnets, reaching $\sim 10$ picoseconds for sub-$kT$ barriers. While this implies a lower retention time for nonvolatile memory applications, as will be discussed later, the faster octupole fluctuations provides a distinct advantage for probabilistic computing applications.

\begin{figure}
    \centering
    \includegraphics[width=\columnwidth]{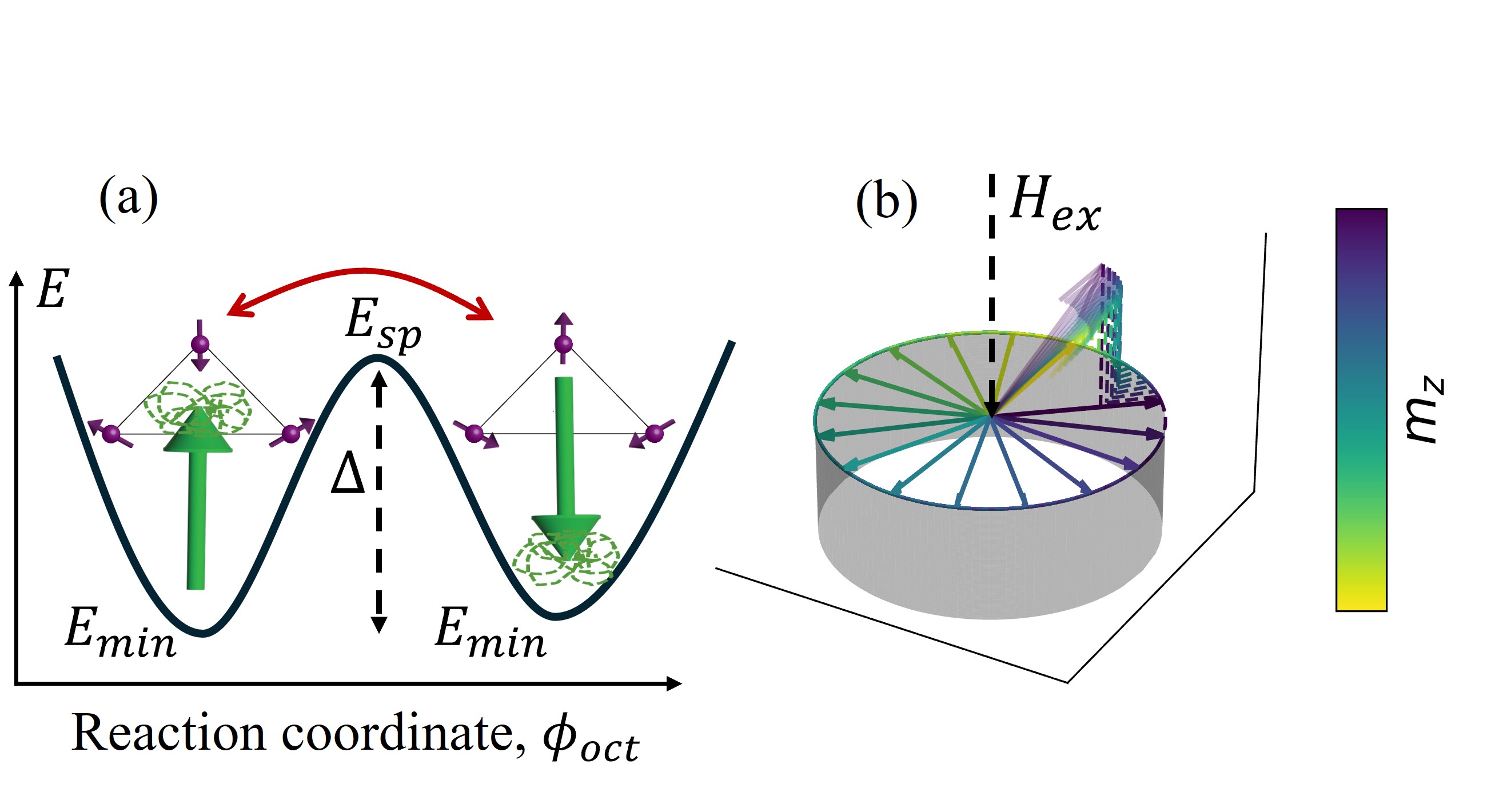}
    \caption{(a) Schematic of the high-barrier limit mechanism of stochastic octupole dynamics. Within an energy minimum, thermal fields primarily induce small angle fluctuations of the green octupole moment. Occasionally, the octupole moment receives a thermal kick large enough to escape over the barrier and precess into the adjacent minimum. (b) Schematic of the low-barrier limit mechanism of octupole fluctuations. Thermal fields induce small excursions ($m_z\ll 1$) of the dipole moment out of the Kagome plane. For brevity, only a few deviations above the plane are shown with their magnitude pictured by the color bar. These out-of-equilibrium moments experience restoring fields ($\propto m_z$) towards the Kagome plane, about which they execute Larmor precessions of varying frequencies. The octupole moments in the kagome plane lose correlations as they dephase relative to each other due to these precessions.}
    \label{fig:3}
\end{figure}

\textit{Analytics.}|We now turn towards gaining an analytical understanding of the underlying relaxation mechanisms and the enhanced rate of octupole fluctuations observed in the simulations. We first focus on the RTN behavior observed in the high-barrier limit. In this case, the simulations show that the octupole moment mostly precesses at small angles around its easy axis in one energy minimum before a random thermal kick quickly rotates it within the easy plane to the other minimum (see Fig.~\ref{fig:3}(a)). Thus, the problem of finding the relaxation time in the high-barrier limit reduces to calculating the mean escape time $\tau_{\text{esc}}$ for octupole rotations within the $xy$ plane. 
Kinetic rate theory based on Langer's formalism \cite{langer_statistical_1969} provides a powerful tool to calculate $\tau_{\rm esc}$ between two energy minima separated by a barrier by using free energy and the deterministic dissipative equations of motion in the system's multi-dimensional phase space. This escape time is given by \cite{kramers_brownian_1940, langer_statistical_1969, hanggi_reaction-rate_1990, coffey_thermal_2012}: $\tau_{\text{esc}}= A^{-1} (\delta F/kT) \tau_{\text{ihd}}$. Note that the octupole relaxation time $\tau = 0.25\tau_{\text{esc}}$ to allow for the loss of correlations as the octupole escapes back and forth over the barrier \cite{coffey_thermal_2012}. Here,
\begin{equation}
    \tau_{\text{ihd}}=\frac{2\pi}{\lambda_+} \frac{V_{\min}}{V_{\text{sp}}} (2 \pi k T)^{\frac{P_{\text{sp}} - P_{\min}}{2}} \sqrt{\frac{\prod_j |\epsilon_{j, \text{sp}}|}{\prod_j \epsilon_{j, \min}}} e^{\Delta/kT} 
    \label{eqn:tau_ihd}
\end{equation}
is the escape time in the so-called intermediate to high damping regime, i.e. when the energy lost ($\delta \mathcal{F}$) due to dissipative coupling to the environment over an equal energy contour containing the energy maxima is larger than the thermal energy $kT$ \cite{kramers_brownian_1940, langer_statistical_1969}. $A(x) = \exp\left[\int_{-\infty}^{\infty} \, dy \ln\left(1 - e^{-x\left(0.25 + y^2\right)}\right) / 2\pi (0.25 + y^2) \right]$ is the dimensionless depopulation factor, which accounts for increase in the escape time over $\tau_{\text{ihd}}$ as the dissipative coupling to the bath is reduced \cite{melnikov_theory_1986}. $\epsilon_{\text{j,min}}$ ($\epsilon_{\text{j,sp}}$) is the curvature obtained from harmonic approximation of system's free energy $\mathcal{F}$ along the $j$th variable at the minima (saddle points). $P_{\text{sp}}$ ($P_{\text{min}}$) denotes the number of Goldstone modes (i.e.  the variables along which free energy variation has a zero curvature) and $V_{\text{sp}}$ ($V_{\text{min}}$) denotes the phase space volume of the corresponding Goldstone modes at energy minima (saddle points). $\lambda_+$ is the positive eigenvalue of the linearized equations of motion around the relevant first-order saddle point \cite{rozsa_reduced_2019}. 

To make analytical progress for deriving $\tau_{\rm esc}$ for octupoles flipping via the path of easy-plane rotations observed in spin dynamics simulations, we exploit the hierarchy of energy scales ($J > D\gg K$) to reduce the 6-dimensional chiral AFM system (Eq.~(\ref{eqn:hamiltonian})) into an effective 2-dimensional system. To this end, following Ref.~\cite{dasgupta_theory_2020}, we first perform a change of basis from $\vec{m}_i$ to normal mode coordinates of the Mn$_3$X system in the exchange limit (i.e. when $D=K=\nu=0$ in Eq.~(\ref{eqn:hamiltonian})) \cite{supplementary}. These modes correspond to the canonically conjugate pairs of net spin canting (in the $x$, $y$ or $z$ directions) and the corresponding rigid rotations (in the $yz$, $xz$ or $xy$ planes, respectively) of the Mn$_3$X spin motif. Note that since rigid rotations in the exchange limit do not cost any energy, these modes are degenerate \textit{zero modes}(see supplementary material). 

The first advantage of this basis is that one of the normal modes --- the mode with net spin canting along $z$ and spin rotations in the $xy$ plane --- directly corresponds to the octupole dynamics of interest to this work, i.e. its rotation within the easy plane (referred here as the $xy$-octupole mode). Secondly, when $D$, $K$ and $\nu$ are turned on, we notice that the degeneracy of the normal modes is lifted with the $xy$-octupole mode acquiring a gap $\sim \sqrt(JK\nu$) while the other two normal modes acquire a gap $\sim \sqrt(JD)$ \cite{supplementary}. Since $K\nu \ll D$, the $xy$-octupole mode remains the lowest energy mode dominating the long-time dynamics. An effective theory can then be constructed for the octupole dynamics of interest by perturbatively expanding the free energy (Eq.~(\ref{eqn:hamiltonian})) up to second order in the small normal mode coordinates and adiabatically eliminating the higher energy modes (i.e. setting them to their $xy$-octupole mode dependent instantaneous equilibrium values). The details of the calculations are presented in the supplementary material \cite{supplementary}. The reduced free energy and dissipative equations of motion governing the octupole dynamics obtained by this procedure are given by \cite{supplementary}:

\begin{equation}
V\mathcal{F}_{\text{oct}} = \frac{3}{2} M_s H_{\text{J}}V m_z^2 + \Delta\sin^2 \phi_{\text{oct}},
\label{eqn:Hoct}
\end{equation}
and 
\begin{equation}
\begin{aligned}
\dot{\phi}_{\text{oct}} - \alpha \dot{m}_z &= \gamma H_{\text{J}} m_z,  \\
\dot{m}_z + \alpha \dot{\phi}_{\text{oct}} &= -\gamma H_{\text{K}} \sin\phi_{\text{oct}} \cos\phi_{\text{oct}}.
\label{eqn:eqofmotion}
\end{aligned}
\end{equation}
Here, $\Delta= {2 \nu K J V }/{(J + \sqrt{3}D)}$ is the octupole energy barrier, $H_{\text{J}}=(3J + \sqrt{3}D)/M_s$ is the strength of the exchange field (note, $J\gg D$), $H_{\text{K}}=2\Delta/3M_sV$ is the strength of the effective uniaxial anisotropy, $m_z=\cos\theta$ is the $z$ component of dipole moment and $ \phi_{\text{oct}}$ is the azimuthal angle of the octupole moment.
These equations map on to the free energy and dynamics of an XY magnet: exchange fields ($\sim 100$ T) in the chiral AFM play the role of the dipole fields ($\sim 1$ T) in XY magnet (forcing $m_z$ to 0 in equilibrium), and weak symmetry breaking within the easy plane gives rise to a uniaxial anisotropy in the plane. 

Applying Langer's theory (Eq.~(\ref{eqn:tau_ihd})) to the reduced 2D subspace of octupole fluctuations using $\mathcal{F}_{\rm oct}$ (Eq.~(\ref{eqn:Hoct})) and the equations of motion (Eq.~(\ref{eqn:eqofmotion})) yields \cite{supplementary}:
\begin{equation}
\tau= 
\frac{\pi \, \exp({\Delta/kT}) \, \cdot A^{-1} \left(V\delta F_\text{oct}/kT\right)}{\gamma H_{\text{J}} 
\left[ \sqrt{\left(\alpha(1 + h_p)\right)^2 + 4h_p} - \alpha(1 - h_p) \right]},
\label{eqn:tau_ihd_final}
\end{equation}
where we define $h_p = H_{\text{K}}/H_{\text{J}}$ as a dimensionless parameter. For typical chiral AFM material parameters, $A^{-1} \left(V\delta F_\text{oct}/kT\right)  =1$ (see supplementary for an analytical expression).

Eq.~(\ref{eqn:tau_ihd_final}) is the first main result of this section and is plotted in Fig.~\ref{fig:2}(e), showing excellent agreement with numerical simulations across all studied parameters. Furthermore, it provides insight into why the octupole fluctuations are sped up over perpendicular anisotropy (PMA) and XY ferromagnets. Namely, in XY magnets, it has recently been established that the presence of additional internal easy-plane dipole fields enhances the dipole fluctuations over PMA \cite{kaiser_subnanosecond_2019, kanai_theory_2021}. Since in chiral AFM the strong easy plane character for octupoles arises from the exchange energies that are at least an order of magnitude larger than the dipole fields in XY, the fluctuation speeds are further enhanced.

Next, we turn to deriving analytical expression for the relaxation time in the low-barrier limit of $\Delta\ll kT$. Here, octupole rotations in the easy plane encounter no sizable barrier. Hence, the escape over barrier mechanism and Langer's theory for $\tau$ are not applicable. Instead, the loss of correlations in this case can be understood through a spin dephasing mechanism, in close analogy with that recently proposed for XY magnets \cite{kaiser_subnanosecond_2019}. To illustrate this, we focus on the no-damping limit (which will be justified post priori) of octupole dynamics in the absence of a thermal barrier. The deterministic equations of motion (Eq.~(\ref{eqn:eqofmotion})) then simplify to:
\begin{equation}
\phi_{\text{oct}} = \gamma H_{\text{J}} m_z t.
\label{eqn:phi_oct_motion}
\end{equation}

Thermal fields induce random fluctuations ($|m_z|\ll 1$) of the dipole moment perpendicular to the Kagome plane, which experiences a restoring exchange field, $H_{\text{J}} m_z$, proportional to the deviation (see Fig.~\ref{fig:3}(b)). The octupole moment then executes Larmor precessions around this restoring field according to Eq.~(\ref{eqn:phi_oct_motion}). The range of $m_z$ excited by the thermal fields is in turn governed by the Boltzmann distribution. Fig.~\ref{fig:3}(b) shows how such excitations experience varying restoring fields and precess at varying rates. As a result, the octupole loses its correlation as the moments dephase relative to each other over time. The loss of octupole correlations due to this dephasing mechanism can be analytically quantified as \cite{jaynes_information_1957, brown_thermal_1963, kaiser_subnanosecond_2019},
\begin{equation}
C(t) = \frac{\int_{-1}^{1} dm_z \, \cos(\gamma H_{\text{J}} m_z t) \, \exp(-\mathcal{F}_{\text{oct}} V/kT)  }{\int_{-1}^{1} dm_z \, \exp(-\mathcal{F}_{\text{oct}} V/kT)},
\label{eqn:tau_autocorr}
\end{equation}
where $C(t)$ is the normalized autocorrelation function of octupole fluctuations (see also Fig.~\ref{fig:2}(d)), which yields $C(t) \approx \text{exp}(- \omega^2_{\text{J}} t^2/2)$, with $\omega_{\text{J}} = \gamma \sqrt{H_{\text{J}} H_{\text{th}}}$ and $H_{\text{th}}=kT/M_sV$ \cite{supplementary}. This gives the low-barrier autocorrelation time,
\begin{equation}
\tau = \frac{1}{\gamma} \sqrt{\frac{2 ln(2)}{H_{\text{J}} H_{\text{th}}}}.
\label{eqn:tau_lowbarrrier}
\end{equation}

Fig.~\ref{fig:2}(e) and Eq.~(\ref{eqn:tau_lowbarrrier}) show excellent agreement between numerical simulations and analytical estimations $\tau$ in the low-barrier limit. These asymptotes are the second main result of this section. In this limit,  the relaxation time is independent of the choice of the Gilbert damping constant $\alpha$, which similar to the easy plane limit, is valid for 
$\alpha \ll \sqrt{H_{th}/H_{\text{J}}}$ \cite{kaiser_subnanosecond_2019}. Furthermore, similar to the high-barrier limit, the correlation time for chiral antiferromagnets is orders of magnitude lower than that estimated for XY ferromagnets due to the exchange-field enhanced ultrafast dynamics in antiferromagnets.

\textit{Electrical tunability.}|Due to the similar mathematical structure of the underlying order parameter and the symmetries of the Hamiltonian, spin dynamics in magnets with a strong XY character has long been known to closely resemble the order parameter dynamics in superfluids \cite{takei_superfluid_2014}. More recently, the possibility of triggering this dynamics via spin current injection has led to the development of various superfluid-inspired spintronic devices \cite{upadhyaya_magnetic_2017, PhysRevB.90.220401, PhysRevB.95.144402, liu_synthetic_2020}. Building on the XY character of chiral antiferromagnets and the experimentally demonstrated ability to trigger octupole dynamics in Mn$_3$X through spin injection \cite{tsai_electrical_2020, takeuchi_chiral-spin_2021}, we next propose and demonstrate a new scheme for electrically tuning the relaxation timescales of chiral AFM. This approach is inspired by the analogy between chiral antiferromagnets subjected to spin currents and current-biased Josephson junctions (JJ). 

\begin{figure}[t]
    \centering
    \includegraphics[width=0.49\textwidth]{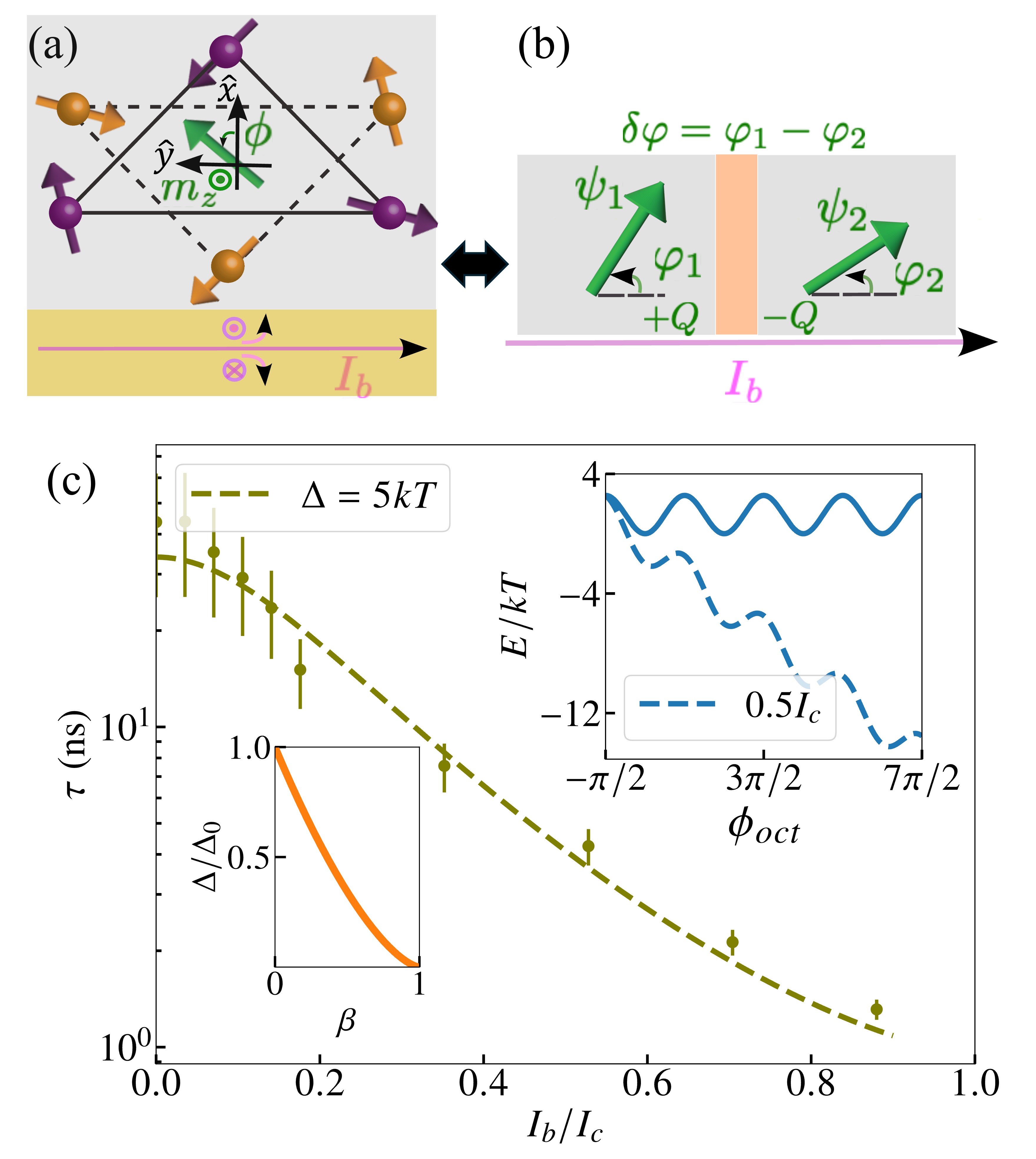}
    \caption{(a) Schematic of the Mn$_3$X nanomagnet coupled to thermal baths and spin injection sources. The nanomagnet is grown with its kagome plane perpendicular to the substrate. The heavy metal layer carrying charge currents below the magnet injects spin currents polarized orthogonal to the kagome plane. (b) Analogy between chiral AFM and a current biased Josephson junction. (c) Numerical simulations (solid dots) and the analytical fit (dashed line) of thermal dynamics of a chiral AFM in the presence of spin currents. The top inset shows the energy landscape of the chiral AFM in the absence (solid line) and in the presence of spin currents (dashed line) depicting the so-called tilted washboard potential. The bottom inset shows this energy barrier as a function of the charge current in the heavy metal layer. }
    \label{fig:4}
\end{figure}

The state of a JJ is described by the canonically conjugate pair variables: $(\delta \varphi, \hbar/2e \,Q)$. Here, $\delta\varphi$ and $Q$ are the difference of order parameter phases across the junction and the (excess) charge on superfluid islands, respectively (see Fig.~\ref{fig:4}(b)); $\hbar$ is the reduced Planck constant and $e$ is the charge of the electron. JJ's free energy can be written as the sum of charging and Josephson's energies as \cite{supplementary} $\mathcal{F}_{\rm JJ}^0= Q^2/2C + 2E_{\text{JJ}} \cos^2(\delta \varphi/2)$, with $C$ and $E_{\text{JJ}}$ being the junction's capacitance and the Josephson energy parameter, respectively. By identifying the canonically conjugate pair $\left(-3M_sVm_z/\gamma, \phi_{\text{oct}} \right)$ with $\left(\hbar Q/2e, \, \delta\varphi \right)$, the first and second terms of the reduced octupole free energy in chiral antiferromagnets (Eq.~(\ref{eqn:Hoct})) can be viewed as analogs of the charging and the Josephson energy, respectively. When the JJ is biased by a charge current $I_b$, $\delta\varphi$ evolves to a new value, which in turn induces a voltage via the Josephson relation $\dot{\delta\varphi} = 2e\mathcal{V}/\hbar$ with $\mathcal{V}=Q/C$. The electrical work done in this process adds a current-dependent term, $\hbar I_b \delta\varphi/2e$, to the junction's free energy, creating a tilted washboard potential energy for the ``$\delta\varphi$-particle": $2E_{\rm JJ} \cos^2 (\delta\varphi/2) + \hbar I\varphi/2e $ \cite{takei_superfluid_2014}. An equivalent term can be generated for a chiral AFM when a spin current polarized perpendicular to the easy plane is injected into the magnet \cite{supplementary} (see Fig.~\ref{fig:4}(a)). In this case, the work done by the spin torque $\partial_t \vec{m}=-\gamma H_{\text{S}}(\vec{m}\times \vec{m} \times \vec{z}$) adds a $-3 M_s H_\text{S}V\phi_{\text{oct}}$ term to the free energy \cite{supplementary, slonczewski_current-driven_1996, sun_spin-current_2000} with $H_\text{S}=\hbar \theta_\text{sh}I_b/2e(3M_s)V$ \cite{supplementary} and $\theta_\text{sh}$ being the spin-Hall angle.
Crucially, this implies that the energy barrier for octupole rotations in the resultant tilted washboard potential can be tuned from a finite value down to 0 as the current is dialed from 0 to $I_c=2e\Delta/\hbar\theta_\text{sh}$ (see insets of Fig.~\ref{fig:4}(c)). Consequently, from Eq.~(\ref{eqn:tau_ihd_final}), the octupole relaxation times can be tuned by orders of magnitude by spin currents.

Mn$_3$X family offers practical advantages to realize the scheme above. First, chiral AFMs can be grown with their Kagome planes oriented orthogonal to the film's growth direction \cite{Yoon_2020}. This allows for the use of charge currents flowing in conventional heavy metals (e.g. Pt, W, Ta) integrated with the magnet to inject the desired spin current via the spin Hall effect \cite{sinova_spin_2015}. Second, due to the weak in-plane anisotropy, the critical currents $I_c$ are small, implying an enhanced sensitivity of relaxation times to the current; indeed experiments have already demonstrated rotation of octupoles in thin films by spin injection \cite{tsai_electrical_2020, takeuchi_chiral-spin_2021}. To test the proposed scheme, we show in Fig.~\ref{fig:4}(c) $\tau$ for an Mn$_3$Sn nanoparticle as a function of biasing currents. This is obtained from (a) numerically integrating the stochastic LLG equations modified to include the spin torque term, and (b) by applying the Langer formalism to the tilted washboard potential (see supplementary), which yields:
\begin{equation}
\tau = \frac{\pi}{\lambda_+} \frac{e^{\Delta_{\uparrow\downarrow}/kT} e^{\Delta_{\downarrow\uparrow}/kT}}{(e^{\Delta_{\uparrow\downarrow}/kT} + e^{\Delta_{\downarrow\uparrow}/kT})}.
\end{equation}
Here, the current dependence mainly arises from electrical tunability of the thermal barriers with $\Delta_{\uparrow\downarrow}=\Delta (\sqrt{1 - \beta^2} + \beta \sin^{-1}(\beta)  -  \pi\beta/2)$ and $\Delta_{\downarrow\uparrow}=\Delta_{\uparrow\downarrow}+\pi\Delta\beta$ with $\beta=I_b/I_c$ being the ratio of bias current to the critical current. $\lambda_{+}$ is the positive eigenvalue of the linearized equations of motion around the energy maximum of the tilted washboard potential (see the supplement for its analytical expression \cite{supplementary}). This is the main result of this section, showing excellent agreement between numerical simulations and our analytical calculations, corroborating the current-biased JJ-inspired tuning of octupole relaxation in chiral antiferromagnets. 

\textit{Outlook.}|In summary, we show that chiral antiferromagnets host up to picosecond scale octupole fluctuations, with electrically tunable relaxation times.  On the theoretical front, we have focused on temperature scales below the ordering temperature, thus neglecting longitudinal fluctuations of octupoles. It would also be interesting to explore them in a future work. On the applications front, recent demonstrations of octupole-driven TMR \cite{chen_octupole-driven_2023, qin_room-temperature_2023, chou_large_2024} combined with our results suggests that chiral antiferromagnets could contribute to the development of emerging technologies requiring the generation of on-chip tunable random numbers \cite{10.1063/5.0067927, 10068500}. For instance, when integrated with transistors, stochastic chiral AFM-based magnetic tunnel junctions could form a probabilistic(p-) bit, the basic building block for the emerging paradigm of probabilistic computing \cite{PhysRevX.7.031014, borders_integer_2019, 10068500, Banerjee2024, Konakanchi2024}. Compared to the current dipole-order-based p-bits, octupole-based p-bits would offer faster fluctuation speeds for the same energy barrier (see Fig.~\ref{fig:2}), reducing time-to-solution \cite{PhysRevApplied.12.034061, aadit_massively_2022, chowdhury_accelerated_2023}. Nanodots of chiral AFM with energy barriers relevant for such p-bit applications have already been experimentally demonstrated \cite{sato_thermal_2023}. Furthermore, due to the negligible dipole moment, octupole fluctuations are expected to be less sensitive to stray magnetic fields, making octupole-based p-bits more robust than their dipole-based counterparts. Additionally, the proposed tunability of correlation times via electric current could be used to dynamically correct for device-to-device variations.
We hope that our results will thus inspire experiments to extend current demonstrations of octupole-based MTJs into the nanoscale fluctuating regime and with improved TMR values, aiming to enable faster and more robust probabilistic computing.

\begin{acknowledgments}
STK and PU would like to thank Supriyo Datta, Kerem Camsari, Jonathan Sun, Kirill Belashchenko and Zhihong Chen for helpful discussions. STK and PU acknowledge support from the National Science Foundation (NSF) grants DMREF-2324203 and ECCS-2331109. YY acknowledges support from JSPS Kakenhi (Grant Nos.~23H01828 and 22KK0072), and JST TI-FRIS. SK and SF acknowledge support from JSPS Kakenhi (Grant Nos.~24H00039 and 24H02235); MEXT X-NICS (Grant No.~JPJ011438); JST-ASPIRE (Grant No.~JPMJAP2322); and JST-CREST (Grant No.~JPMJCR19K3).
\end{acknowledgments}

\appendix

\newpage
\onecolumngrid

\section{\label{sec:suppmicro}Micromagnetic Simulations}

\begin{figure*}[b]
    \centering
    \includegraphics[width=0.67\textwidth]{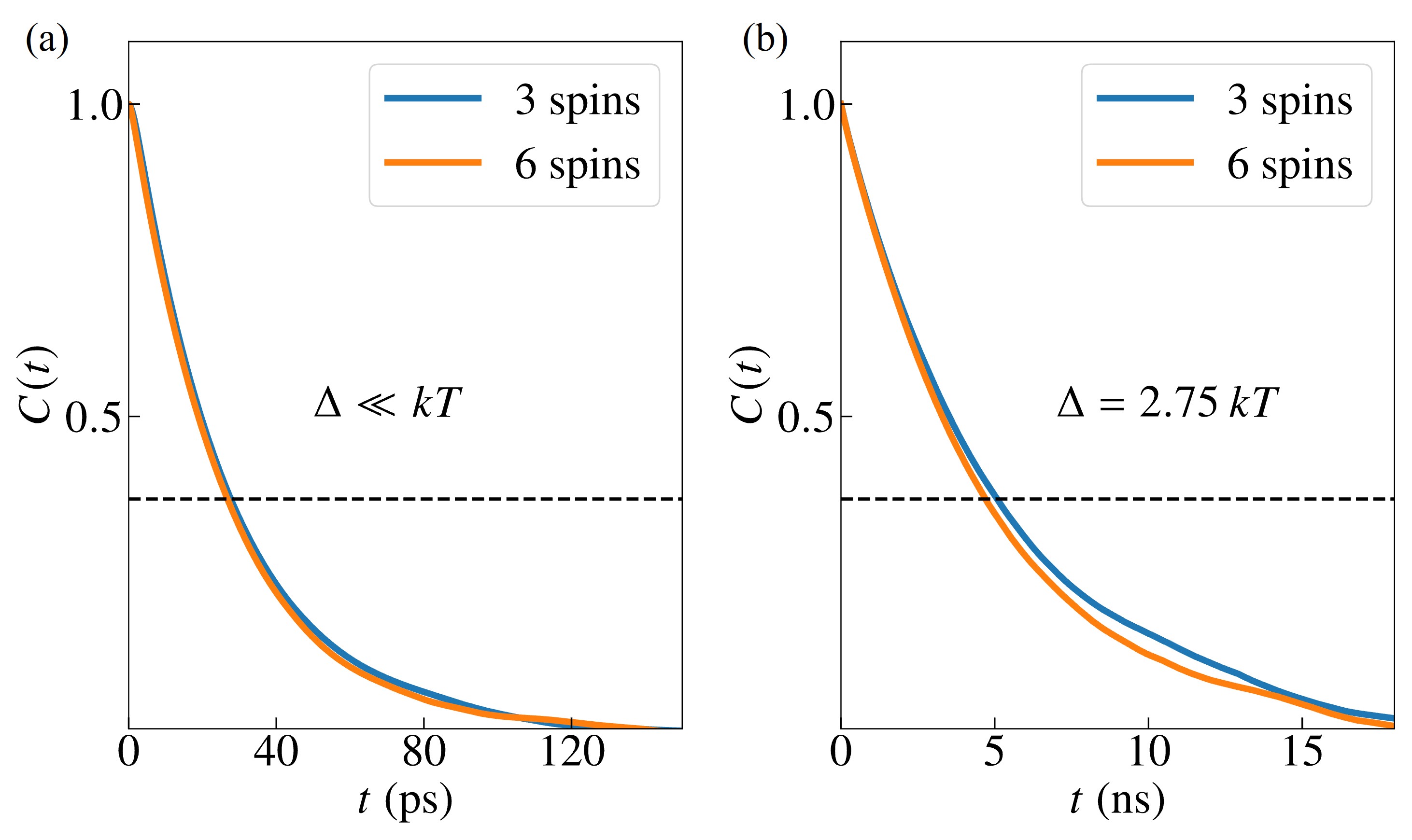}
    \caption{Autocorrelation function plots of the stochastic octupole moment in the strong easy-plane limit (along its easy axis) for 3-spin and 6-spin limits of the free energy. We compare the approaches for a range of thermal barriers from the low-barrier limit to the high-barrier limit. The dashed line shows $1/e$ value of the autocorrelation function.}
    \label{supp:six-spin}
\end{figure*}

The unit cell of the Mn$_3$X system consists of 6 Mn spins arranged in two stacked kagome planes (in a star motif) as highlighted in the main text. There, we focused on the dynamical modes where each Mn spin in one Kagome plane is ferromagnetically locked with its inverted neighbor in the adjacent kagome plane. Here, we justify our approach of neglecting the out-of-phase dynamics within such inter-plane pairs of Mn spins in the experimentally relevant time and energy scales. To this end, we study the stochastic dynamics of the Mn$_3$X system in the monodomain continuum limit for both 3-spin and 6-spin configurations. Such a monodomain limit is valid when the size of the system is smaller than the micromagnetic exchange length, which is the case for nanomagnets. Including strain, the free energy of the system in the three-spin limit is given by:
\begin{equation}
\begin{aligned}
\mathcal{F} = & \, J \left( \vec{m}_1 \cdot \vec{m}_2 + \vec{m}_2 \cdot \vec{m}_3 + \vec{m}_3 \cdot \vec{m}_1 \right) - J \nu \left(\vec{m}_1 \cdot \vec{m}_2 \right) + D \left[ \hat{z} \cdot \left( \vec{m}_1 \times \vec{m}_2 + \vec{m}_2 \times \vec{m}_3 + \vec{m}_3 \times \vec{m}_1 \right) \right] \\
& - K \left[ \left( \vec{e}_1 \cdot \vec{m}_1 \right)^2 + \left( \vec{e}_2 \cdot \vec{m}_2 \right)^2 + \left( \vec{e}_3 \cdot \vec{m}_3 \right)^2 \right]. 
\\ 
\label{eq:supphamiltonian}
\end{aligned}
\end{equation}
Here, $\vec{M}_i = M_s \vec{m}_i$ represent three sublattice magnets of saturation magnetization $M_s=400$ emu/cc. Each sublattice magnet corresponds to a pair of ferromagnetically locked inter-plane Mn spins. $\vec{e}_1 = -\sqrt{3}/{2} \, \hat{x} -{1}/{2}  \, \hat{y}$, $\vec{e}_2 = -\sqrt{3}/{2} \, \hat{x} +{1}/{2}  \, \hat{y}$ and $\vec{e}_3 =\hat{y}$ are the uniaxial anisotropy directions for the three magnets \cite{tsai_electrical_2020, higo_perpendicular_2022, takeuchi_chiral-spin_2021}. In the 6-spin configuration, we introduce three more sublattice magnets (the index $i$ now runs to 6) and reduce each sublattice saturation magnetization to $M_s=200$ emu/cc to keep the same total saturation magnetization in both cases. In the second kagome plane, the magnets $\vec{m}_4$, $\vec{m}_5$ and $\vec{m}_6$ have a free energy density analogous to Eq.~\ref{eq:supphamiltonian}, with $\vec{e}_4=\vec{e}_1$, $\vec{e}_5=\vec{e}_2$ and $\vec{e}_6=\vec{e}_3$ being the respective easy axes. Additionally, we add the term $\sum_{i=1}^3 -J\vec{m}_i \cdot \vec{m}_{i+3}$ to the total free energy to account for the net ferromagnetic coupling between the inter-plane spins \cite{liu_anomalous_2017}.

 For both 3-spin and 6-spin configurations, we study the thermal dynamics of the strained Mn$_3$X system by numerically integrating the stochastic Landau–Lifshitz–Gilbert equation (sLLG):
 \begin{equation}
     \partial_t \vec{m}_i=-\gamma \vec{m}_i \times (-\delta_{M_s\vec{m}_i} \mathcal{F}+ \vec{H}^{\rm th}_i) + \alpha \vec{m}_i \times \partial_t \vec{m}_i.
     \label{eq:suppsLLG}
\end{equation}
Here, $\gamma$ is the magnitude of the electron's gyromagnetic ratio and $\vec{H}^{\rm th}_i$ are thermal fields that satisfy the fluctuation dissipation theorem \cite{brown_thermal_1963, 1060329} -- $\langle H^{\rm th}_i(t) \rangle=0$ and $\langle H^{\rm th}_i(t) H^{\rm th}_j(t') \rangle= \sigma^2\delta_{ij}\delta(t-t')$ with $\sigma^2=2 \alpha k T/\gamma M_s V$, where $\delta$ is the Dirac delta function. The simulations are performed using a homegrown HSPICE based compact circuit models for the chiral AFM \cite{camsari_modular_2015}. As a model candidate for the Mn$_3$X family with anti-chiral ground state, we perform all numerical analyses for Mn$_3$Sn with the parameters \cite{tsai_electrical_2020, higo_perpendicular_2022, takeuchi_chiral-spin_2021, liu_anomalous_2017, yamane_dynamics_2019, shukla_spin-torque-driven_2022, sato_thermal_2023}: $J=59\times 10^6$ J/m$^3$, $D=0.1 J$, $K=110\times 10^3$ J/m$^3$ and $\nu=0.0006$.

Fig.~\ref{supp:six-spin} shows a close agreement between the autocorrelation functions of the stochastic octupole moment in the 3-spin and the 6-spin configurations. The difference in the relaxation times obtained from the two approaches is within the margin of error for a range of barriers from the low-barrier limit to the high-barrier limit. This indicates that the pairs of Mn spins on the opposite ends of the Mn$_3$X star motif are locked ferromagnetically within the time and energy scales of interest. So throughout the main text, we focus on the 3-spin configuration for its analytical and computational simplicity.

\section{Two Coordinate Effective Free Energy}
\subsection{Normal Modes}

\begin{figure*}[t]
    \centering
    \includegraphics[width=\textwidth]{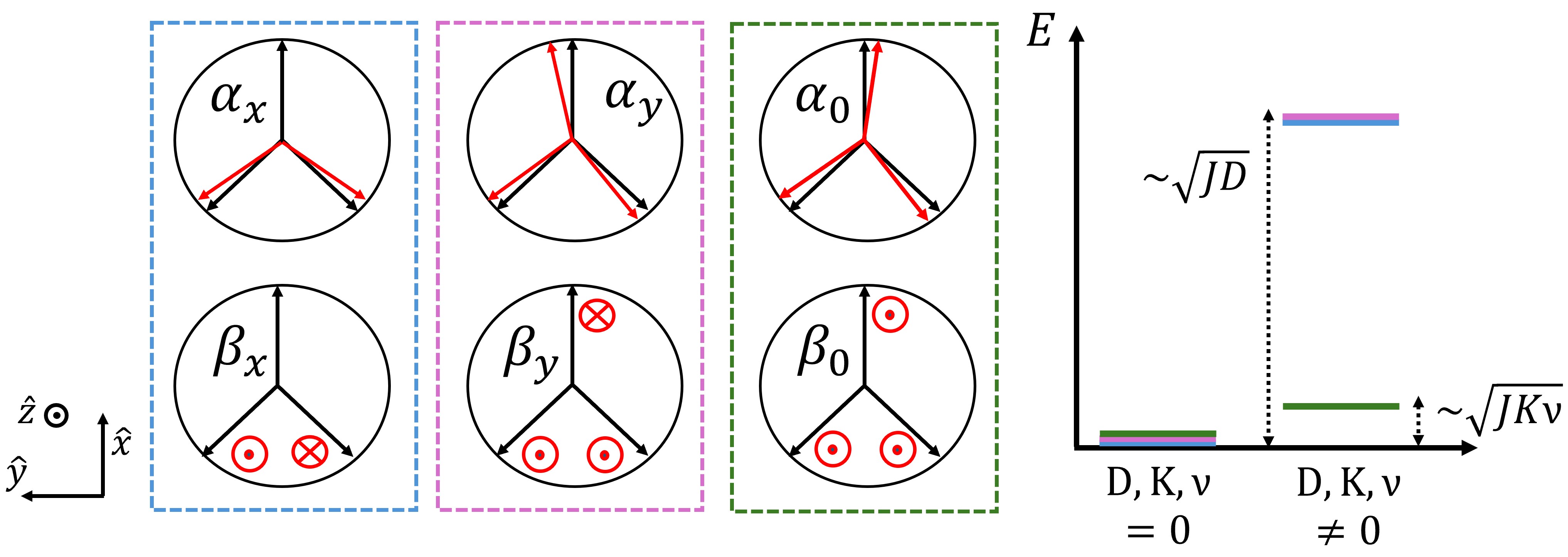}
    \caption{Schematic representation of the normal modes of the Mn$_3$X system in the exchange limit \cite{dasgupta_theory_2020}. In this case, the normal modes are the degenerate zero energy modes of rotations of the Mn$_3$X motif about the $x$, $y$ and $z$ axes. When $D, K$ and $\nu$ are turned on, the normal modes acquire a gap as shown in the figure with the $xy$-octupole mode still being the lowest energy, and hence the most relevant mode for stochastic dynamics.}
    \label{fig:suppnormalmodes}
\end{figure*}

Having justified our approach of using the effective 3-spin free energy, we exploit the hierarchy of energy scales ($J > D\gg K$) to further reduce this 6-dimensional free energy (Eq.~\ref{eq:supphamiltonian}) into an effective 2-dimensional free energy for the octupole dynamics of interest. To this end, following Ref.~\cite{dasgupta_theory_2020}, we perform a change of basis from $\vec{m}_i=(\sin\theta_i\cos\phi_i, \sin\theta_i\sin\phi_i,  \cos\theta_i)$ to normal mode coordinates of the Mn$_3$X system in the exchange limit (i.e., when $D=K=\nu=0$). As shown in Fig.~\ref{fig:suppnormalmodes}, these modes correspond to the canonically conjugate pairs of spin canting ($\alpha_x$, $\alpha_y$ and $\beta_0$) and the corresponding rigid rotations ($\beta_x$, $\beta_y$ and $\alpha_0$) of the Mn$_3$X spin motif about the $x$, $y$ and $z$ axes, respectively. Note that in the exchange limit, rigid rotations about all three axes are degenerate zero energy modes as highlighted in Fig.~\ref{fig:suppnormalmodes}. The basis vectors of these normal modes can be related to ($\theta_i$, $\phi_i$) as \cite{dasgupta_theory_2020}:
\begin{equation}
\begin{pmatrix}
\phi_1 \\
\phi_2 \\
\phi_3
\end{pmatrix}
= \begin{pmatrix}
2\pi/3 \\
4\pi/3 \\
0
\end{pmatrix}
+ L \begin{pmatrix}
\alpha_x \\
\alpha_y \\
\alpha_0
\end{pmatrix},
\label{eq:3}
\end{equation}

\begin{equation}
\begin{pmatrix}
\theta_1 \\
\theta_2 \\
\theta_3
\end{pmatrix}
= \begin{pmatrix}
\pi/2 \\
\pi/2 \\
\pi/2
\end{pmatrix}
+L \begin{pmatrix}
\beta_x \\
\beta_y \\
\beta_0
\end{pmatrix},
\label{eq:4}
\end{equation}

with
\noindent
\begin{equation}
L = \begin{pmatrix}
1/\sqrt{2} & 1/\sqrt{6} & 1/\sqrt{3} \\
-1/\sqrt{2} & 1/\sqrt{6} & 1/\sqrt{3} \\ 
0 & -2/\sqrt{6} & 1/\sqrt{3}
\end{pmatrix}.
\end{equation}
Here, the columns of the orthogonal matrix \(L\) are obtained by applying a rigid rotation ($R_x$, $R_y$ or $R_z$) on the Mn$_3$X spin motif (see Fig.~\ref{fig:suppnormalmodes}). The first advantage of this basis is that one of the modes ($\alpha_0, \beta_0$) directly corresponds to the octupole dynamics of interest. Secondly, when $D$, $K$ and $\nu$ are turned on, we notice that the degeneracy of the normal modes is lifted, with the $xy$-octupole mode ($\alpha_0$ mode) acquiring a gap $\sim \sqrt(JK\nu$) while the other two normal modes acquire a gap $\sim \sqrt(JD)$ \cite{supplementary}. Since $K\nu \ll D$, the $xy$-octupole mode remains the lowest energy mode dominating the long-time dynamics. Furthermore, note that $\alpha_x$, $\alpha_y$, $\beta_x$, $\beta_y$ and $\beta_0$ incur either exchange or DMI energy, while $\alpha_0$ incurs weaker anisotropy energy. An effective theory can thus be constructed for the octupole dynamics by perturbatively expanding the free energy (Eq.~\ref{eq:supphamiltonian}) up to second order in the small normal mode coordinates ($\alpha_x$, $\alpha_y$, $\beta_x$, $\beta_y$ and $\beta_0$) and adiabatically eliminating the high energy modes. 
\subsection{Second Order Perturbation Theory}
In this subsection, we perturbatively expand our free energy in the small normal mode coordinates. In the strong easy-plane limit, as outlined in the main text, the octupole moment may be given as,
\begin{equation}
    \vec{m} = \frac{1}{3}M_{xz} \left( \vec{m}_3 + R \vec{m}_1 + R^2 \vec{m}_2 \right).
    \label{eqn:octupole_supp}
\end{equation}
Here $R$ represents a $\frac{2\pi}{3}$ rotation operation about the $z$-axis. Using Eqs.~\ref{eq:3}--\ref{eqn:octupole_supp}, the octuple moment $\vec{m}=(\sin \theta\cos\phi_{\text{oct}}, \\ \sin\theta\sin\phi_{\text{oct}}, \cos\theta)$ can be further expressed in the normal mode coordinates as (note that the octupole moment transforms like a pseudovector under the mirror operator):
\begin{equation}
\begin{aligned}
\sin \phi_{\text{oct}} &=  \sin \left(\frac{\alpha_0}{\sqrt{3}}\right),
\\
- \cos \theta &= \sin \left(\frac{\beta_0}{\sqrt{3}}\right).
\label{eq:octupole_angles}
\end{aligned}
\end{equation}

\noindent\textbf{Exchange Interaction (J)}:
We employ a second-order perturbative approach \cite{he_magnetic_2024} to expand $\mathcal{F}$ in terms of normal modes by substituting Eq.~\ref{eq:3} and Eq.~\ref{eq:4} and ignoring higher order terms. We start with the exchange energy term.
\begin{align}
\mathcal{F} &= \mathcal{F}_{J_{\textit{a}}} + \mathcal{F}_{J_{\textit{b}}} + \mathcal{F}_{J_{\textit{c}}} \notag \\
&= \quad J \big[ \sin\theta_1 \sin\theta_2 \cos(\phi_1 - \phi_2) + \cos\theta_1 \cos\theta_2 \big] \notag \\
&\quad + J \big[ \sin\theta_2 \sin\theta_3 \cos(\phi_2 - \phi_3) + \cos\theta_2 \cos\theta_3 \big] \notag \\
&\quad + J \big[ \sin\theta_3 \sin\theta_1 \cos(\phi_3 - \phi_1) + \cos\theta_3 \cos\theta_1 \big].
\end{align}

\begin{equation}
\begin{aligned}
\mathcal{F}_{J_{\textit{a}}} = & \, J \left[ \sin \left( \frac{\pi}{2} + \frac{\beta_x}{\sqrt{2}} + \frac{\beta_y}{\sqrt{6}} + \frac{\beta_0}{\sqrt{3}} \right) \sin \left( \frac{\pi}{2} - \frac{\beta_x}{\sqrt{2}} + \frac{\beta_y}{\sqrt{6}} + \frac{\beta_0}{\sqrt{3}} \right) \cos \left( \frac{2\pi}{3} + \sqrt{2}\alpha_x \right) \right. \\
& \left. + \cos \left( \frac{\pi}{2} + \frac{\beta_x}{\sqrt{2}} + \frac{\beta_y}{\sqrt{6}} + \frac{\beta_0}{\sqrt{3}} \right) \cos \left( \frac{\pi}{2} - \frac{\beta_x}{\sqrt{2}} + \frac{\beta_y}{\sqrt{6}} + \frac{\beta_0}{\sqrt{3}} \right) \right],
\\
\mathcal{F}_{J_{\textit{b}}} = & \, J \left[ \sin \left( \frac{\pi}{2} - \frac{\beta_x}{\sqrt{2}} + \frac{\beta_y}{\sqrt{6}} + \frac{\beta_0}{\sqrt{3}} \right) \sin \left( \frac{\pi}{2} - \frac{2\beta_y}{\sqrt{6}} + \frac{\beta_0}{\sqrt{3}} \right) \cos \left( \frac{2\pi}{3} - \frac{\alpha_x}{\sqrt{2}} + \frac{3}{\sqrt{6}} \alpha_y \right) \right. \\
& \left. + \cos \left( \frac{\pi}{2} - \frac{\beta_x}{\sqrt{2}} + \frac{\beta_y}{\sqrt{6}} + \frac{\beta_0}{\sqrt{3}} \right) \cos \left( \frac{\pi}{2} - \frac{2\beta_y}{\sqrt{6}} + \frac{\beta_0}{\sqrt{3}} \right) \right],
\\
\mathcal{F}_{J_{\textit{c}}} = & \, J \left[ \sin \left( \frac{\pi}{2} - \frac{2 \beta_y}{\sqrt{6}} + \frac{\beta_0}{\sqrt{3}} \right) \sin \left( \frac{\pi}{2} + \frac{\beta_x}{\sqrt{2}} + \frac{\beta_y}{\sqrt{6}} + \frac{\beta_0}{\sqrt{3}} \right) \cos \left( -\frac{4\pi}{3} - \frac{\alpha_x}{\sqrt{2}} - \frac{3 \alpha_y}{\sqrt{6}} \right) \right. \\
& \left. + \cos \left( \frac{\pi}{2} - \frac{2 \beta_y}{\sqrt{6}} + \frac{\beta_0}{\sqrt{3}} \right) \cos \left( \frac{\pi}{2} + \frac{\beta_x}{\sqrt{2}} + \frac{\beta_y}{\sqrt{6}} + \frac{\beta_0}{\sqrt{3}} \right) \right].
\label{eq:Jexpansion}
\end{aligned}
\end{equation}

Expanding $\mathcal{F}$ up to second order in $\alpha_x$, $\alpha_y$, $\beta_x$, $\beta_y$ and $\beta_0$, we get,
\begin{equation}
\mathcal{F}_J = J \left[ -\frac{9}{2}\left(1-\frac{\beta_0^2}{3}\right) + \frac{3}{4} \left( \alpha_x^2 + \alpha_y^2 \right) \right].
\end{equation}
\\
\noindent\textbf{Strain ($\nu$)}: The strain term is given by $\mathcal{F}_{\nu} = - J\nu \left(\vec{m_1} \cdot \vec{m_2}\right)$, with $\nu<<1$. Expanding this term to second order gives,
\begin{equation}
   \mathcal{F}_{\nu} = J \nu \sqrt{\frac{3}{2}} \alpha_x.
\end{equation}

\noindent\textbf{Dzyaloshinskii–Moriya Interaction (D)}: We similarly expand the DMI term up to second order.
\begin{align}
\mathcal{F}_D &= \mathcal{F}_{D_a} + \mathcal{F}_{D_b} + \mathcal{F}_{D_c} \notag \\
&= D \sin\theta_1 \sin\theta_2 \sin(\phi_2 - \phi_1) \notag  + D \sin\theta_2 \sin\theta_3 \sin(\phi_3 - \phi_2) \notag + D \sin\theta_3 \sin\theta_1 \sin(\phi_1 - \phi_3).
\end{align}
\begin{equation}
\begin{aligned}
\mathcal{F}_{D_a} = & \,  D \sin \left( \frac{\pi}{2} + \frac{\beta_x}{\sqrt{2}} + \frac{\beta_y}{\sqrt{6}} + \frac{\beta_0}{\sqrt{3}} \right) \sin \left( \frac{\pi}{2} - \frac{\beta_x}{\sqrt{2}} + \frac{\beta_y}{\sqrt{6}} + \frac{\beta_0}{\sqrt{3}} \right) \sin \left(-\frac{2\pi}{3} - \sqrt{2} \alpha_x \right),
\\
\mathcal{F}_{D_b} = & \,  D \sin \left( \frac{\pi}{2} - \frac{\beta_x}{\sqrt{2}} + \frac{\beta_y}{\sqrt{6}} + \frac{\beta_0}{\sqrt{3}} \right) \sin \left( \frac{\pi}{2} - \frac{2 \beta_y}{\sqrt{6}} + \frac{\beta_0}{\sqrt{3}} \right) \sin \left( -\frac{2\pi}{3} + \frac{\alpha_x}{\sqrt{2}} - \frac{3 \alpha_y}{\sqrt{6}} \right), \\
\mathcal{F}_{D_c} = & \,  D \sin \left( \frac{\pi}{2} - \frac{2 \beta_y}{\sqrt{6}} + \frac{\beta_0}{\sqrt{3}} \right) \sin \left( \frac{\pi}{2} + \frac{\beta_x}{\sqrt{2}} + \frac{\beta_y}{\sqrt{6}} + \frac{\beta_0}{\sqrt{3}} \right)  \sin \left( \frac{4\pi}{3} + \frac{\alpha_x}{\sqrt{2}} + \frac{3 \alpha_y}{\sqrt{6}} \right).
\end{aligned}
\end{equation}

\begin{equation}
\mathcal{F}_{D} = \frac{\sqrt{3} D}{2} \left( -3 \left(1-\frac{\beta_0^2}{3}\right) + \beta_x^2 + \beta_y^2 \right) + \frac{3 \sqrt{3} D}{4} \left( \alpha_x^2 + \alpha_y^2 \right).
\end{equation}

\noindent\textbf{Uniaxial Anisotropy (K)}
We expand the uniaxial anisotropy up to second order in the normal modes as,
\begin{equation}
\begin{aligned}
\mathcal{F}_K = & \, \mathcal{F}_{K_a} + \mathcal{F}_{K_b} + \mathcal{F}_{K_c} \\
= & \, -K \sin^2 \theta_1 \sin^2 \left(\frac{\pi}{3} + \phi_1 \right) -K \sin^2 \theta_2 \cos^2 \left( \frac{\pi}{3} - \phi_2 \right) -K \sin^2 \theta_3 \sin^2 \phi_3.
\end{aligned}
\end{equation}

\begin{equation}
\begin{aligned}
\mathcal{F}_{K_a} = & \, -K \sin^2 \left( \frac{\pi}{2} + \frac{\beta_y}{\sqrt{2}} + \frac{\beta_x}{\sqrt{6}} + \frac{\beta_0}{\sqrt{3}} \right) \sin^2 \left( \frac{5\pi}{3} + \frac{\alpha_x}{\sqrt{2}} + \frac{\alpha_y}{\sqrt{6}} + \frac{\alpha_0}{\sqrt{3}} \right),
\\
\mathcal{F}_{K_b} = & \, -K \sin^2 \left( \frac{\pi}{2} - \frac{\beta_y}{\sqrt{2}} + \frac{\beta_x}{\sqrt{6}} + \frac{\beta_0}{\sqrt{3}} \right)  \sin^2 \left( -\frac{\pi}{3} - \frac{\alpha_x}{\sqrt{2}} + \frac{\alpha_y}{\sqrt{6}} + \frac{\alpha_0}{\sqrt{3}} \right),
\\
\mathcal{F}_{K_c} = & \, -K \sin^2 \left( \frac{\pi}{2} - \frac{2 \beta_x}{\sqrt{6}} + \frac{\beta_0}{\sqrt{3}} \right)  \sin^2 \left( - \frac{2 \alpha_y}{\sqrt{6}} + \frac{\alpha_0}{\sqrt{3}} \right).
\end{aligned}
\end{equation}

\begin{equation}
\mathcal{F}_K = K \sqrt{\frac{3}{2}} \left[ \alpha_x \cos \left( \frac{2 \alpha_0}{\sqrt{3}} \right) + \alpha_y \sin \left( \frac{2 \alpha_0}{\sqrt{3}} \right) \right].
\end{equation}

\noindent\textbf{Total Free Energy:} The total free energy then is, $\mathcal{F} = \mathcal{F}_J + \mathcal{F}_{\nu} + \mathcal{F}_{D} + \mathcal{F}_K$.

\subsection{Adiabatic Elimination}

The hierarchy of energy scales $J>D \gg K \sim J\nu$ results in a separation of timescales with the high frequency modes ($\alpha_x$, $\alpha_y$, $\beta_x$, $\beta_y$) reaching steady-state much faster than the slow modes  ($\alpha_0$ and $\beta_0$). This enables us to employ adiabatic elimination and obtain an effective one-mode Hamiltonian predominantly described in terms of $\alpha_0$ and $\beta_0$.

\begin{align}
\left. \frac{\partial \mathcal{F}}{\partial \beta_x} \right|_{\beta_x = \beta_{x,eq}} &= \left. \frac{\partial \mathcal{F}}{\partial \beta_y} \right|_{\beta_y = \beta_{y,eq}} = 0 \implies \beta_{x,eq} = \beta_{y,eq} = 0 \\
\left. \frac{\partial \mathcal{F}}{\partial \alpha_x} \right|_{\alpha_x = \alpha_{x,eq}} &= 0
\implies \alpha_{x,eq} = \frac{-J \nu - K \cos \left( \frac{2 \alpha_0}{\sqrt{3}} \right)}{\sqrt{\frac{3}{2}} (J + \sqrt{3} D)} \\
\left. \frac{\partial \mathcal{F}}{\partial \alpha_y} \right|_{\alpha_y = \alpha_{y,eq}} &= 0
\implies \alpha_{y,eq} = - \frac{ K \sin \left( \frac{2 \alpha_0}{\sqrt{3}} \right)}{\sqrt{\frac{3}{2}} (J + \sqrt{3} D)}.
\end{align}

Substituting the magnitudes of $\alpha_{x,eq}$, $\alpha_{y,eq}$, $\beta_{x,eq}$ and $\beta_{y,eq}$ into $\mathcal{F}$, we obtain the final free energy density in terms of the normal mode coordinates $\alpha_0$ and $\beta_0$ as:
\begin{equation}
\mathcal{F}_{\text{oct}} = \left[\frac{9J + 3\sqrt{3}D}{2} \left(\frac{\beta_0}{\sqrt{3}}\right)^2 + \frac{2 J K \nu}{J +  \sqrt{3}D} \sin^2 \left( \frac{\alpha_0}{\sqrt{3}} \right)\right].
\label{eq:finalH_normalMode}
\end{equation}

Substituting $\alpha_0$ and $\beta_0$ from Eq.~\ref{eq:octupole_angles} in Eq.~\ref{eq:finalH_normalMode} and noting that $m_z = \beta_0/\sqrt{3}$, we obtain the desired effective low-energy Hamiltonian,
\begin{equation}
V\mathcal{F}_{\text{oct}} = \frac{3}{2} M_s H_{\text{J}}V m_z^2 + \Delta\sin^2 \phi_{\text{oct}},
\label{eq:suppfinalH}
\end{equation}
where, $\Delta= {2 \nu K J V }/{(J + \sqrt{3}D)}$ is the octupole energy barrier and $H_{\text{J}}=(9J + 3\sqrt{3}D)/3M_s$ is the strength of the exchange field (note, $J\gg D$).

\section{Octupole Equations of Motion}

\noindent\textbf{Conservative equation of motion}: For an electron, the net angular momentum in the $z$ direction is the generator of rotations, $\phi_{\text{e}}$, in the $xy$ plane. Thus, we identify $-3VM_sm_{z,e}/\gamma$ and $\phi_{\text{e}}$ as canonically conjugate variables obeying Hamilton's equations of motion. Here, $\gamma>0$ is the magnitude of the electron's gyromagnetic ratio. For the free energy in Eq.~\ref{eq:suppfinalH}, the conservative equations of motion are,

\begin{equation}
\begin{aligned}
\dot{\phi_{\text{e}}} &=  -\frac{\gamma}{3M_s}\left(9J + 3\sqrt{3}D\right) m_{z,\text{e}} = -\gamma H_{\text{J}} m_{z,\text{e}}, \\
\dot{m}_{z,\text{e}} &= \frac{\gamma}{3 M_s} \left( \frac{2 J K \nu}{ J + \sqrt{3}D}  \right) \sin \left(2\phi_{\text{e}}\right) = \gamma H_{\text{K}}\sin \phi_{\text{e}} \cos \phi_{\text{e}}.
\end{aligned}
\end{equation}
Here, $H_{\text{K}}=2\Delta/3M_sV$. From Eq. \ref{eqn:octupole_supp}, with the mirror operation on the Mn spins, the conservative equations of motion for the octupole can be written as,
\begin{equation}
\begin{aligned}
\dot{\phi_{\text{oct}}} &= \gamma H_{\text{J}} m_z, \\
\dot{m_z} &= -\gamma H_{\text{K}}\sin \phi_{\text{oct}} \cos \phi_{\text{oct}}.
\end{aligned}
\end{equation}

\noindent\textbf{Deterministic Dissipative Equations of motion}: Including Gilbert damping, the linearized LLG equation for an electron is given by,
\begin{equation}
\begin{aligned}
\dot{\theta_\text{e}} &= \frac{\gamma}{1 + \alpha^2} \left( 
    - H_{\text{K}} \sin\theta_\text{e} \sin\phi_\text{e} \cos\phi_\text{e} 
    + \alpha H_{\text{J}} \sin\theta_\text{e} \cos\theta_\text{e} 
    + \alpha H_{\text{K}} \sin\theta_\text{e} \cos\theta_\text{e} \cos^2\phi_\text{e}
\right),
\\
\dot{\phi_\text{e}} &= \frac{\gamma}{1 + \alpha^2} \left(
    - H_{\text{J}} \sin\theta_\text{e} \cos\theta_\text{e} 
    - H_{\text{K}} \sin\theta_\text{e} \cos\theta_\text{e} \cos^2\phi_\text{e} 
    - \alpha H_{\text{K}} \sin\theta_\text{e} \sin\phi_\text{e} \cos\phi_\text{e}
\right).
\end{aligned}
\end{equation}

Here too, the mirror operation in Eq. \ref{eqn:octupole_supp} changes the octupole equations of motion to

\begin{equation}
\begin{aligned}
\dot{\theta} &= -\frac{\gamma}{1 + \alpha^2} \left( 
    - H_{\text{K}} \sin\theta \sin\phi_{\text{oct}} \cos\phi_{\text{oct}} 
    - \alpha H_{\text{J}} \sin\theta \cos\theta 
    - \alpha H_{\text{K}} \sin\theta \cos\theta \cos^2\phi_{\text{oct}}
\right),
\\
\dot{\phi}_{\text{oct}} &= -\frac{\gamma}{1 + \alpha^2} \left(
    - H_{\text{J}} \sin\theta \cos\theta 
    - H_{\text{K}} \sin\theta \cos\theta \cos^2\phi_{\text{oct}} 
    + \alpha H_{\text{K}} \sin\theta \sin\phi_{\text{oct}} \cos\phi_{\text{oct}}
\right).
\label{eq:octupoleEOM}
\end{aligned}
\end{equation}

\section{Derivation of The Relaxation Time in High Barrier Limit}
We employ Langer's theory on the effective octupole free energy $\mathcal{F}_{\text{oct}}$ to derive an analytical formula for the octupole relaxation time in the high barrier limit of $\Delta > kT$. As highlighted in the main text, thermal fields mostly induce small angle precessions of the octupole moment around its easy axis in an energy minimum. Once in a while, the octupole moment receives a large thermal kick that quickly rotates it within the easy plane to the adjacent minimum. Within Langer's theory, the mean escape time from one well to the other is given by \cite{kramers_brownian_1940, langer_statistical_1969, hanggi_reaction-rate_1990, coffey_thermal_2012-1},
\begin{equation}
\tau_{\text{esc}}= A^{-1} \left(\frac{V\delta \mathcal{F}}{kT}\right) \tau_{\text{ihd}}.
 \label{eq:supplanger}
\end{equation}

In Langer's theory, in the so-called intermediate to high damping (IHD) regime \cite{kramers_brownian_1940, langer_statistical_1969, coffey_crossover_2001, IHD2, IHD3, IHD4}, a fluctuating particle in a metastable energy well is assumed to quickly attain and maintain a Maxwell-Boltzmann distribution of energies within the well \cite{desplat_thermal_2019}. A Gibbs ensemble of such particles allows for a small fraction of the particles near the saddle point energy, $E_{sp}$, to escape quasi-statically over the barrier to the adjacent minimum \cite{duff_langers_2008}. This intermediate to high damping limit escape time is given by escape time is given by \cite{kramers_brownian_1940, langer_statistical_1969, hanggi_reaction-rate_1990, coffey_thermal_2012-1}:
\begin{equation}
    \tau_{\text{ihd}}=\frac{2\pi}{\lambda_+} \frac{V_{\min}}{V_{\text{sp}}} (2 \pi k T)^{\frac{P_{\text{sp}} - P_{\min}}{2}} \sqrt{\frac{\prod_j |\epsilon_{j, \text{sp}}|}{\prod_j \epsilon_{j, \min}}} e^{\Delta/kT}. 
    \label{eqn:supptau_ihd}
\end{equation}
In this formula, the eigenvalues $\epsilon_{\text{j,min}}$ and $\epsilon_{\text{j,{\text{sp}}}}$ are obtained from the harmonic approximation of the octupole free energy near the energy minima and the saddle points, respectively. $\lambda_+$ is the positive eigenvalue of the linearized Landau-Lifshitz-Gilbert equation of motion around the saddle point. To account for Goldstone modes in our system, a factor of $2 \pi k T$ is introduced; $P$ denotes the number of Goldstone modes and $V$ represents the phase-space volume corresponding to the Goldstone modes.

In the so-called very low damping (VLD) limit, i.e. when the energy lost ($V\delta \mathcal{F}$) due to dissipative coupling to the environment over an equal energy contour containing the energy maxima is smaller than the thermal energy $kT$, the octupole energies within an energy minimum deviate from the Maxwell-Boltzmann distribution of the IHD limit. To account for these deviations and the reduced dissipative coupling to the thermal bath, a depopulation factor $A(V\delta \mathcal{F}/kT)$ is added on to $\tau_{\text{ihd}}$ \cite{depop2, coffey_crossover_2001, rozsa_reduced_2019} with
\begin{equation}
A(x) = \exp\left\{\frac{1}{2\pi} \int_{-\infty}^{\infty} \ln\left[1 - e^{-x\left(\frac{1}{4} + y^2\right)}\right] \frac{1}{\frac{1}{4} + y^2} \, dy\right\}.
\end{equation}

Therefore, the escape time $\tau_{\text{esc}}$ in the most general case is then given by Eq.~\ref{eq:supplanger}. First, we note that the octupole relaxation time $\tau = 0.25\tau_{\text{esc}}$ to allow for the loss of correlations as the octupole escapes back and forth over the barrier \cite{coffey_thermal_2012}. Second, as we shall see in the following subsections, the depopulation factor evaluates to unity for typical Mn$_3$Sn parameters, i.e. the magnet is in the IHD limit. However, with the very low Gilbert damping constants being reported more recently in some of the chiral AFMs in the Mn$_3$X family, the depopulation factor may deviate significantly from unity (note, $A^{-1}(x) \geq1,  \forall x$).

\subsection{Calculation of intermediate-to-high damping relaxation time}
\noindent\textbf{Energy Minima and Saddle Point}: At the minima, $\theta=\pi/2$ and $\phi_{\text{oct}}=(0,\pi)$ such that $\mathcal{F}_{min}=0$. Similarly, at the saddle point, $\theta=\pi/2$ and $\phi_{\text{oct}}=(\pi/2,3\pi/2)$ gives $\mathcal{F}_{\text{{\text{sp}}}}=\Delta=\frac{2 J K \nu V}{\left(J + \sqrt{3} D\right)}$.\\

\noindent\textbf{Calculation of Eigenvalues and Goldstone Modes}: 
A Taylor series expansion around the minimum and saddle points helps us obtain the eigenvalues of the harmonic approximation of the Hamiltonian.
\begin{align}
\epsilon_{1,\text{{\text{sp}}}} &= V\frac{\partial^2 \mathcal{F}_{\text{oct}}}{\partial m_z^2} \Bigg|_{m_z = 0, \phi_{\text{oct}} = \pi/2} = 3 M_\text{s} H_\text{J} V, \\
\epsilon_{2,\text{{\text{sp}}}} &= V\frac{\partial^2 \mathcal{F}_{\text{oct}}}{\partial \phi_{\text{oct}}^2} \Bigg|_{m_z = 0, \phi_{\text{oct}} = \pi/2} = -2 \Delta, \\
\epsilon_{1,\text{min}} &= V\frac{\partial^2 \mathcal{F}_{\text{oct}}}{\partial m_z^2} \Bigg|_{m_z = 0, \phi_{\text{oct}} = \pi} = 3 M_\text{s} H_\text{J} V, \\
\epsilon_{2,\text{min}} &= V\frac{\partial^2 \mathcal{F}_{\text{oct}}}{\partial \phi_{\text{oct}}^2} \Bigg|_{m_z = 0, \phi_{\text{oct}} = \pi} = 2 \Delta,
\end{align}
\begin{equation}
\sqrt{\frac{\prod_j |\epsilon_{j, \text{{\text{sp}}}}|}{\prod_j \epsilon_{j, \min}} } = 1.
\end{equation}
We further have, $P_{\text{{\text{sp}}}} = P_{\text{min}} = 0$ \cite{goldstonemodes} and
\begin{align}
\frac{V_{\text{min}}}{V_{\text{{\text{sp}}}}} \left( 2 \pi k_B T \right)^{\frac{P_{\text{{\text{sp}}}} - P_{\text{min}}}{2}} = 1.
\end{align} \\

$\lambda_+$ is the positive eigenvalue of the linearized Landau–Lifshitz–Gilbert equation of motion around the saddle point ($\theta=\pi/2$ and $\phi_{\text{oct}}=\pi/2$). Now, linearizing the equations of motion (Eq. \ref{eq:octupoleEOM}) at the saddle point gives,
\begin{equation}
\begin{pmatrix}
\delta \dot{\theta} \\
\delta \dot{\phi_{\text{oct}}}
\end{pmatrix}
= -\frac{\gamma}{1 + \alpha^2}H_{\text{J}}
\begin{pmatrix}
\alpha & h_p \\
1 & -\alpha h_p
\end{pmatrix}
\begin{pmatrix}
\delta \theta \\
\delta \phi_{\text{oct}}
\end{pmatrix},
\end{equation}
Here, $h_p = H_\text{K} / H_\text{J}$ is a dimensionless constant. The positive eigenvalue $\lambda_+$ is then given by,
\begin{equation}
\lambda_+ = \frac{\gamma}{(1 + \alpha^2)}\cdot H_{\text{J}} 
\left[
\frac{-\alpha(1 - h_p) + \sqrt{\left(\alpha(1 + h_p)\right)^2 + 4h_p}}{2}
\right].
\label{lambda_+}
\end{equation}
Finally, the Langer formula gives,
\begin{equation}
\tau_{\text{ihd}} =
\frac{4 \pi}{\gamma H_\text{J} 
\left[ \sqrt{\left(\alpha(1 + h_p)\right)^2 + 4h_p} - \alpha(1 - h_p) \right]} 
\exp\left(\frac{\Delta}{k T}\right).
\end{equation}

\subsection{Calculation of Depopulation Factor (A)}
To evaluate $A^{-1} \left( V{\delta\mathcal{F}_\text{oct}}/{k T} \right)$, we begin by evaluating $ V\delta\mathcal{F}_{\text{oct}}$, the energy dissipated in a single precession of the octupole moment along an equal-energy contour containing the saddle points. The dissipated energy can be obtained from the damping term of the LLG-like equation for the octupole moment: $ \partial_t \vec{m}=-\alpha \vec{m} \times \partial_t \vec{m}$. This gives,
\begin{equation}
\delta \mathcal{F}_{\text{oct}} = \int_0^{t} \left( \frac{\partial \mathcal{F}_{\text{oct}}}{\partial m_z} \dot m_{z}|_\text{diss} \, dt + \frac{\partial \mathcal{F}_{\text{oct}}}{\partial \phi_{\text{oct}}} \dot{\phi}_{\text{oct}}|_{\text{diss}} \, dt \right).
\label{eq:depop_energy}
\end{equation}
To evaluate $\dot{m}_{z}|_\text{diss}$ and $\dot{\phi}_{\text{oct}}|_{\text{diss}}$ we can solve for,
\begin{equation}
   \frac{d\vec{m}}{dt} \Bigg|_{\text{diss}} = -\alpha \left( \vec{m} \times \frac{d\vec{m}}{dt} \right), 
\end{equation}
with
$\vec{m} = \left( \cos\phi_{\text{oct}}\sqrt{1-m_z^2} , \sin\phi_{\text{oct}}\sqrt{1-m_z^2}, m_z \right)$
and
\[
\frac{d\vec{m}}{dt} = \left( - \dot{\phi}_{\text{oct}} \sin \phi_{\text{oct}}, \dot{\phi}_{\text{oct}} \cos \phi_{\text{oct}}, \dot{m_z}
\right).
\]
This gives $\dot{m}_{z}|_\text{diss} = -\alpha \dot{\phi}_{\text{oct}}$ and $\dot{\phi}_{\text{oct}}|_{\text{diss}} =  \alpha \dot{m}_{z}$.
Therefore, 
Eq. \ref{eq:depop_energy} can be re-written as,
\begin{equation}
\delta \mathcal{F}_{\text{oct}} =  \int_{\phi_{\text{oct}}} \left( \frac{\partial \mathcal{F}_\text{oct}}{\partial m_z} \right)(-\alpha) \, d\phi_{\text{oct}} 
+ \int_{m_z} \left( \frac{\partial \mathcal{F}_\text{oct}}{\partial \phi_{\text{oct}}} \right) (\alpha) \, d{m_z}.
\end{equation}\label{eq:depop_energy_v2}
Note that due to the coupled nature of these integrals, we need to obtain relationships between $m_z$ and $\phi_{\text{oct}}$. Since the equal-energy contour must satisfy $\mathcal{F}_{\text{oct}} = \Delta$, so we can use the following expression to perform the integrals over their respective variables.

\begin{equation}
    m_z = \sqrt{\frac{2 \Delta}{3 M_\text{s} H_\textbf{J} V}} \cos \phi_{\text{oct}}.
\end{equation}

We evaluate the above integrals and via appropriate substitutions and algebra, we arrive at the following.
\begin{equation}
V\delta \mathcal{F}_\text{oct} = \alpha \sqrt{6 \Delta M_\text{s} H_\text{J}V} \approx 6 \alpha \sqrt{J K \nu}V.
\end{equation}

Since $x$ = $V\delta\mathcal{F}/k T$, the depopulation factor $A(x)$ could be obtained numerically from the following expression.

\begin{equation}
A(x) = \exp\left\{\frac{1}{2\pi} \int_{-\infty}^{\infty} \ln\left[1 - e^{-x\left(\frac{1}{4} + y^2\right)}\right] \frac{1}{\frac{1}{4} + y^2} \, dy\right\},
\end{equation}
Since arriving at an analytical solution for the above integral is difficult, a numerical approach is employed to calculate the depopulation factor $A(x)$.  However, we notice that for $x \gg 1$, $A(x)\approx1$. In our case, for Mn$_3$Sn parameters considered in this work, $A(x)$ indeed evaluates to unity.

\section{Spin Torque Driven Chiral AFM as a Current-Biased Josephson junction}
In this section, we draw an analogy between a current-biased Josephson junction (JJ) and a spin-torque-driven chiral antiferromagnet (AFM) in the geometry where a spin current polarized perpendicular to the easy plane is injected into the chiral AFM. To establish this analogy, we first briefly review the free energy and the equations of motion for a current-biased JJ \cite{10.1063/1.5089550}.

A Josephson Junction (JJ) is formed by interfacing two superconducting islands with a weak link (either a non or weakly superconducting region). The generalized coordinates describing the state of a 
JJ are given by the canonically conjugate variables: $(\delta \varphi, \hbar/2e \,Q)$. Here, $Q$ is the excess charge on the islands and $\delta \varphi \equiv \varphi_1-\varphi_2$, with  $\varphi_i$ being the phase of the condensed wave function (order parameter) describing Cooper pairs in the $i$th superconducting island. The free energy of JJ can be written as:
\begin{equation}
\mathcal{F}_{\rm JJ}^0= Q^2/2C + 2E_{\text{JJ}} \cos^2\delta \varphi/2.
\end{equation}
Here, the first term is the capacitive charging energy with $C$ being the capacitance of the JJ. The second term is the Josephson energy, which arises from the coupling between the wave functions of the two islands, and is parameterized by the coupling parameter $E_{\text{JJ}}$. The equations of motion for the JJ, as obtained by using the canonical conjugacy between $n$ and $\delta\varphi$, can be written as: $\delta\dot {\varphi}=2e/\hbar \, \partial_Q \mathcal {F}_{\rm JJ}^0$, and $\hbar/2e \,\dot {Q}= -\,\partial_{\delta\varphi} \mathcal {F}_{\rm JJ}^0$, which respectively yields the Josephson's relations: 
\begin{equation}
\dot {\delta\varphi}=\frac{2e\mathcal{V}}{\hbar}; I=I^{\rm JJ}_c \sin \delta \varphi.
\label{eq:suppJJrelations}
\end{equation}
Here $\mathcal{V}=Q/C$ is the voltage across the islands and $I^{\rm JJ}_c=2eE_{\text{JJ}}/\hbar$.

In the absence of an external current, the equilibrium state of the JJ, as obtained by minimizing $\mathcal {F}_{\rm JJ}^0$, is given by $\delta \varphi=0$. When JJ is biased by an external current, $I_{\rm b}$, $\delta \varphi$ evolves to a new equilibrium value. During the transient, the change in $\delta \varphi$ induces a voltage across the JJ according to the Josephson relations (Eq.~\ref{eq:suppJJrelations}). The $I\mathcal{V}$ work done during the transient can be added as an additional contribution to the junction's free energy: $\int \hbar/2e \, I_{b} \dot{\delta\varphi} dt = \hbar/2e \,I_b\delta \varphi$. The junction's free energy in the presence of a current bias is thus given by:
\begin{equation}
\mathcal{F}_{\rm JJ}= Q^2/2C + 2E_{\text{JJ}} \cos^2(\delta\varphi/2)+ \hbar/2e \, I_b \delta \varphi.
\end{equation}
The $\varphi-$dependent part of the above energy is the well-known tilted washboard potential \cite{tinkham2004introduction}. The behavior of JJ under a current bias can be understood from this free energy as follows. As $I_b$ is increased from zero, $\delta\varphi$ evolves to a new equilibrium value, which is given by the local minimum of the titled washboard potential (see Fig.~4 in the main text). Additionally, the barrier between the minima is decreased with increasing $I_b$. Above a critical value $I_b=I^{\rm JJ}_c$, the barrier and the local minima disappear giving rise to the so-called dissipative voltage-state,  with oscillating  $\delta \varphi$ and hence a finite voltage across the junction. 

In direct analogy with above description of JJ we now proceed to write down the free energy and equation of motion of chiral AFM in the presence of spin currents. As highlighted above, the low-energy dynamics of chiral AFM  is described by the easy-plane octupole order parameter. The state of chiral AFM is described by the canonically conjugate pair: $(m_z, \phi_{\text{oct}})$, with the free energy in the absence of spin current given by Eq.~\ref{eq:suppfinalH}. If we identify the canonically conjugate pair: $\left(-3M_sVm_z/\gamma, \phi_{\text{oct}} \right)$ with $\left(\hbar Q/2e, \, \delta\varphi \right)$, the first (second) term above can be thought of as the equivalent to the charging (Josephson) energy.

In the absence of spin current the equilibrium state of the chiral AFM is given by $\phi_{\text{oct}}=0$. When a spin current polarized orthogonal to the easy plane (i.e. along $z$) is injected into the chiral AFM, it applies a torque of the form $\partial_t \vec{m_i}=-\gamma H_{\text{S}}/3 \, \, (\vec{m_i}\times \vec{m_i} \times \vec{p}$) on each of the three sublattice moments, with $\vec{p}$ being the direction of polarization of the spin current. Here, $H_\text{S}= \hbar \theta_{\text{sh}} I_b/2e(3M_s)V$, $ \theta_{\text{sh}}$ is the spin-Hall angle and $I_b$ is the bias charge current in the heavy metal layers. The effective torque exerted on the octupole moment is then given by, $\partial_t \vec{m}=-\gamma H_{\text{S}} \, \, (\vec{m}\times \vec{m} \times \vec{p}$). This effective torque would evolve $\phi_{\text{oct}}$ to a new equilibrium (much like current bias changes $\delta \varphi$ in JJ). The corresponding work done during the transient can be added as an additional free energy term, similar to the $I\mathcal{V}$ work for JJ. This term can be calculated as,
\begin{equation}
\int_0^{t} \left( \frac{\partial \mathcal{F}_{\text{oct}}}{\partial m_z} \dot m_{z}|_\text{ST} \, dt + \frac{\partial \mathcal{F}_{\text{oct}}}{\partial \phi_{\text{oct}}} \dot{\phi}_{\text{oct}}|_{\text{ST}} \, dt \right),
\label{eq:st_energy}
\end{equation}
where, $\dot m_{z}|_\text{ST}$ and $\dot{\phi}_{\text{oct}}|_{\text{ST}}$ are obtained from the octupole spin-torque term. The modified free energy for a chiral AFM in the presence of perpendicular spin injection thus maps onto that of a current biased JJ: 
\begin{equation}
V\mathcal{F}_{oct} = \frac{3}{2} M_s H_{\text{J}}V m_z^2 + \Delta\sin^2 \phi_{\text{oct}} -3M_sH_\text{S}V\phi_{\text{oct}}.
\label{eq:spinH}
\end{equation}
We highlight that the spin current gives rise to a tilted washboard potential for the $xy$-octupole mode (i.e. the $\phi_{\text{oct}}-$dependent term above). Thus, much like the current bias for JJ, the spin current tunes the equilibrium $\phi_{\text{oct}}$ and the energy barrier for octupole to rotate within the easy plane. In particular, the latter forms the basis of the proposed scheme to electrically tune correlation times for octupole fluctuations, as presented in the main text. We also note in passing that above a critical spin current $I_c$ (calculated in the next section), the barrier and local minima disappear. Consequently, in the absence of thermal fluctuations, a static solution  for $\phi_{\text{oct}}$ does not exist for $I> I_c$; instead $\phi_{\text{oct}}$ becomes oscillatory,  analogous to the ``voltage-state'' for JJ. While this state is not useful for probabilistic computing (as it will start adding oscillatory long time correlations to bits), such phenomena can be used to create chiral AFM-based oscillators, as proposed in Ref. \cite{shukla_spin-torque-driven_2022}.  

\section{Barrier Tunability via Spin Current Injection}
To include the effect of spin currents in our Mn$_3$X system, we add the work done by the spin currents as an additional quasi-equilibrium term to the octupole free energy as described in Eq.~\ref{eq:spinH} of the previous section with,
\begin{equation}
H_S = \frac{\hbar}{2e} \cdot \frac{I_b\theta_{\text{sh}}}{3 M_s V}.
\label{eq:spin_curr}
\end{equation}
Here, $\hbar$ is the Planck constant, $I_b$ is the charge current in the heavy metal layer, $\theta_{\text{sh}}$ is the spin-Hall angle, and $V$ is the volume of the chiral AFM. The addition of the work done by the spin currents to the free energy changes the octupole minima ($\phi_{\text{oct}}=0, \pi$) and saddle points ($\pi/2, 3\pi/2$) to

\begin{align}
\phi_1 &=  \frac{1}{2} \sin^{-1} \beta \\
\phi_2 &=  \frac{\pi}{2} - \frac{1}{2} \sin^{-1} \beta \\
\phi_3 &=  \pi + \frac{1}{2} \sin^{-1} \beta \\
\phi_4 &= \frac{3\pi}{2} - \frac{1}{2} \sin^{-1} \beta
\end{align}
Here, $\beta=2H_{\text{S}}/H_{\text{K}}$. Since the equations above do not have solutions if $\beta > 1$, the critical current for lowering the octupole thermal barrier to zero is given by,
\begin{equation}
    I_c = \frac{2JK\nu}{(J +\sqrt{3}D)} \frac{V}{(\hbar/2e) \theta_{\text{sh}}}.
\end{equation}
Beyond this critical current, the spin currents induce coherent oscillations of the octupole moment. The energies of the new minima ($E_1$, $E_3$) and the saddle points ($E_2$, $E_4$) are given by,
\begin{align}
E_1 &= \Delta \left( -\frac{\sqrt{1-\beta^2}}{2} - \frac{\beta}{2} \sin^{-1} \beta \right), \\
E_2 &= \Delta \left( \frac{\sqrt{1-\beta^2}}{2} + \frac{\beta}{2} \sin^{-1} \beta - \frac{\pi }{2}\beta \right), \\
E_3 &= \Delta \left(- \frac{\sqrt{1-\beta^2}}{2} - \frac{\beta}{2} \sin^{-1} \beta - \pi \beta \right), \\
E_4 &= \Delta \left( \frac{\sqrt{1-\beta^2}}{2} + \frac{\beta}{2} \sin^{-1} \beta - \frac{3\pi }{2}\beta \right).
\end{align}
The new energy landscape results in the so-called tilted washboard potential as highlighted in the main text. We now identify two distinct barriers for the two different directions of rotation (clockwise, counterclockwise) of the octupole moment within the easy plane as,
\begin{align}
\Delta_{\uparrow\downarrow} &= (E_2 - E_1) = \Delta\left( \sqrt{1 - \beta^2} + \beta \sin^{-1}(\beta)  - \frac{\pi}{2} \beta  \right),\\
\Delta_{\downarrow\uparrow} &= (E_2 - E_3) = \Delta\left( \sqrt{1 - \beta^2} + \beta \sin^{-1}(\beta)  + \frac{\pi}{2} \beta \right).
\end{align}
Due to the asymmetry in the energy barriers, the octupole moment escapes between the energy minima preferentially by crossing $\Delta_{\uparrow\downarrow}$ rather than by crossing $\Delta_{\downarrow\uparrow}$. These two paths result in two different escape rates, with the net octupole moment relaxation rate being the sum of the two. Thus,
\begin{equation}
    \frac{1}{\tau} = \frac{1}{2} \left(\frac{1}{\tau_{\uparrow\downarrow}}+\frac{1}{\tau_{\downarrow\uparrow}}\right),
\end{equation}
where ${\tau_{\uparrow\downarrow}}$ (${\tau_{\downarrow\uparrow}}$) is the escape time over the barrier ${\Delta_{\uparrow\downarrow}}$ (${\Delta_{\downarrow\uparrow}}$). Next, we use Langer's theory to calculate these two escape times. 

\noindent\textbf{Langer's theory in the presence of spin currents:}

Strictly speaking, as highlighted in the main text, Langer's theory is only valid in the high-barrier limit. So, the following approach is valid when the applied spin current doesn't reduce the barriers below the thermal energy, i.e., $({\Delta_{\uparrow\downarrow}}, {\Delta_{\downarrow\uparrow}}) \geq kT$. Next, we must linearize the LLG around the saddle point and obtain the positive eigenvalue $\lambda_{+}$ in the presence of spin currents. For simplicity, we define
\begin{align}
a &=   \alpha + \frac{\alpha h_p}{2}(1 - \sqrt{1 - \beta^2}), \\
b &= \alpha h_p  \sqrt{1 - \beta^2}, \\
c &=  h_p \left( 1 + \frac{h p}{2} \right) \sqrt{1 - \beta^2} \left( 1 - \sqrt{1 - \beta^2} \right),
\end{align}
and
\begin{equation}
\lambda_+ = \frac{\gamma}{1 + \alpha^2} \cdot H_{\text{J}} \cdot \left(\frac{a + b + \sqrt{(a - b)^2 + 4c}}{2}\right).
\end{equation}
This gives,
\begin{align}
\tau_{\uparrow\downarrow} & = \frac{2\pi}{\lambda_+} e^{\Delta_{\downarrow\uparrow}/kT}\\
\tau_{\downarrow\uparrow} & = \frac{2\pi}{\lambda_+} e^{\Delta_{\uparrow\downarrow}/kT}
\end{align}

\noindent\textbf{Calculation of Depopulation Factor (A):} As previously highlighted, depopulation factors $A_{\rm min}(x)$ and $A_{\rm max}(x)$ can be calculated numerically. For the Mn$_3$Sn parameters used in this text, even in the presence of the spin currents, $A_{\rm min}(x)\approx1$.

\bibliographystyle{apsrev4-2}
\bibliography{apssamp.bib}

\end{document}